\documentclass[twocolumn]{aastex701}

\usepackage{multirow}

\usepackage{amsmath}

\definecolor{Hiro}{rgb}{0,0,1}
\definecolor{GF}{rgb}{1,0,1}
\definecolor{KM}{rgb}{1,0,0}

\begin{document}

\title{The Many Faces of Type Ia Supernova Remnants inside Planetary Nebulae: Morphological Diversity and the Adiabatic Origin of Recombining Plasma}

\author[0009-0007-8699-8973]{Yusei Fujimaru}
\affiliation{Department of Astronomy, Kyoto University, Kitashirakawa, Oiwake-cho, Sakyo-ku, Kyoto 606-8502, Japan}
\email{fujimaru@kusastro.kyoto-u.ac.jp}

\author[0000-0002-2899-4241]{Shiu-Hang Lee}
\affiliation{Department of Astronomy, Kyoto University, Kitashirakawa, Oiwake-cho, Sakyo-ku, Kyoto 606-8502, Japan}
\affiliation{Kavli Institute for the Physics and Mathematics of the Universe (WPI), The University of Tokyo, Kashiwa 277-8583, Japan}
\email{herman@kusastro.kyoto-u.ac.jp}

\author[0000-0002-4231-8717]{Gilles Ferrand}
\affiliation{{The University of Manitoba, Department of Physics and Astronomy, Winnipeg, Manitoba, R3T~2N2, Canada}}
\affiliation{RIKEN Center for Interdisciplinary Theoretical and Mathematical Sciences (iTHEMS), Wak\={o}, Saitama, 351-0198 Japan}
\email{Gilles.Ferrand@umanitoba.ca}

\author[0000-0003-3494-343X]{Carles Badenes}
\affiliation{Department of Physics and Astronomy, University of Pittsburgh, 3941 O’Hara Street, Pittsburgh, PA 15260, USA}
\affiliation{Pittsburgh Particle Physics, Astrophysics, and Cosmology Center (PITT PACC), University of Pittsburgh, Pittsburgh, PA 15260, USA}
\email{badenes@pitt.edu}

\author[0000-0003-2611-7269]{Keiichi Maeda}
\affiliation{Department of Astronomy, Kyoto University, Kitashirakawa, Oiwake-cho, Sakyo-ku, Kyoto 606-8502, Japan}
\email{keiichi.maeda@kusastro.kyoto-u.ac.jp}

\author[0000-0002-7025-284X]{Shigehiro Nagataki}
\affiliation{Astrophysical Big Bang Laboratory (ABBL), RIKEN Pioneering Research Institute (PRI), Wak\={o}, Saitama, 351-0198 Japan}
\affiliation{RIKEN-Berkeley Center, RIKEN iTHEMS, Wak\={o}, Saitama, 351-0198 Japan}
\affiliation{RIKEN Center for Interdisciplinary Theoretical and Mathematical Sciences (iTHEMS), Wak\={o}, Saitama, 351-0198 Japan}
\email{shigehiro.nagataki@riken.jp}

\author[0000-0003-3308-2420]{R\"udiger Pakmor}
\affiliation{Max-Planck-Institut f\"{u}r Astrophysik, Karl-Schwarzschild-Str. 1, D-85748, Garching, Germany}
\email{rpakmor@mpa-garching.mpg.de}

\author[0000-0002-7507-8115]{Daniel Patnaude}
\affil{Smithsonian Astrophysical Observatory, 60 Garden Street, Cambridge, MA 02138 USA}
\email{dpatnaude@cfa.harvard.edu}

\author[0000-0001-6189-7665]{Samar Safi-Harb}
\affiliation{{The University of Manitoba, Department of Physics and Astronomy, Winnipeg, Manitoba, R3T~2N2, Canada}}
\email{samar.safi-harb@umanitoba.ca}

\author[0000-0002-4460-0097]{Friedrich K.\ R{\"o}pke}
\affiliation{Zentrum f{\"u}r Astronomie der Universit{\"a}t Heidelberg, Astronomisches Rechen-Institut, M{\"o}nchhofstr.\ 12--14, 69120 Heidelberg, Germany}
\affiliation{Zentrum f{\"u}r Astronomie der Universit{\"a}t Heidelberg, Institut f{\"u}r Theoretische Astrophysik, Philosophenweg 12, 60120 Heidelberg, Germany}
\affiliation{Heidelberger Institut f{\"u}r Theoretische Studien, Schloss-Wolfsbrunnenweg 35, 69118 Heidelberg, Germany}
\email{friedrich.roepke@h-its.org}

\author[0000-0002-1796-758X]{Anne Decourchelle}
\affiliation{Université Paris-Saclay, Université Paris Cité, CEA, CNRS, AIM, 91191, Gif-sur-Yvette, France}
\email{anne.decourchelle@cea.fr}

\author[0000-0002-5044-2988]{Ivo R. Seitenzahl}
\affiliation{Research School of Astronomy and Astrophysics, Australian National University, Canberra, ACT 2611, Australia}
\affiliation{Mathematical Sciences Institute, Australian National University, Canberra, ACT 0200, Australia}
\email{ivoseitenzahl@gmail.com}

\begin{abstract}
Several Type Ia supernova remnants (SNRs) show asymmetric morphologies, relatively high ionization state, and signatures of overionized plasma.
It remains unclear whether these properties originate from the Type Ia supernova (SN) explosion itself or from the interaction with the surrounding environment.
In this work, we investigate the scenario in which a non-spherical planetary nebula (PN) is formed by a common-envelope (CE) mass ejection, and subsequently an SN Ia explodes inside the PN. 
We calculate the 3D evolution of Ia SNRs inside simplified PN-like circumstellar medium (CSM) models taking into account non-equilibrium ionization, and compute synthetic X-ray spectra.
By changing the delay time between the CE ejection and the SN explosion, we vary the characteristic size and density of the CSM and examine how the PN-shaped environment manifests itself in the properties of SNRs. 
We find that the CE-to-SN delay time produces a wide variety of SNR morphologies, plasma states, and X-ray properties.
Shorter delay times lead to stronger ejecta--CSM interaction, producing more asymmetric bipolar morphologies, ring-like ejecta structures, higher ionization states, and clearer signatures of overionized plasma.
These effects also appear in the synthetic X-ray spectra through line emission, continuum features, and Doppler signatures.
Our results show that a single Type Ia explosion model can produce diverse SNR appearances through interaction with PN-like CSM.
The resulting spectra enable direct comparison with high-energy-resolution XRISM/Resolve observations through line centroids, line profiles, Doppler shifts, and recombination features, providing a way to constrain both the explosion model and the CSM environment of Ia SNRs.
\end{abstract}


\section{Introduction} \label{sec:intro}

Type Ia supernovae (SNe Ia) are widely thought to be thermonuclear explosions of white dwarfs (WDs) in binary systems, and have played a central role as standardizable candles in cosmological distance measurements \citep{riess1998observational,perlmutter1999measurements}.
Despite their importance, their progenitor systems and explosion mechanisms remain unclear.
Several binary pathways have been proposed, including single-degenerate (SD) \citep{whelan1973binaries,nomoto1982accreting} and double-degenerate (DD) \citep{Iben1984Supernovaemass,webbink1984double} scenarios, and the explosion may occur either near the Chandrasekhar mass through accretion or at sub-Chandrasekhar mass through helium-shell detonation, merger, or dynamical interaction.
Because binary evolution, including common-envelope (CE) evolution, can shape the circumstellar environment before the explosion, the diversity of SNe Ia may reflect not only the explosion physics but also the pre-explosion environment \citep[for recent reviews, see][]{liu2023type,ruiter2025type}.

Some SNe Ia also show direct evidence for dense circumstellar interaction.
In particular, SNe Ia-CSM exhibit Type Ia-like spectral features together with narrow hydrogen emission lines, indicating interaction with material lost before the explosion \citep{dilday2012ptf,silverman2013type,jerkstrand2020type, uno2023sn,sharma2023systematic}.
Although rare, accounting for only $\sim0.3\%$ of SNe Ia in the ZTF DR2
sample \citep{dimitriadis2025ztf}, these events demonstrate that thermonuclear explosions can occur in complex CSM rather than only in a simple uniform interstellar medium (ISM).
A recent analysis of the ASAS-SN sample further supports the rarity of SNe Ia-CSM, estimating a relative volumetric rate of $0.036\pm0.017\%$ \citep{desai2026supernova}.

A similar diversity is seen in the remnant phase.
Young Type Ia SNRs preserve the ejecta--environment interaction history, which can be probed through multi-wavelength observations, especially thermal X-ray emission \citep[e.g., the Kepler SNR showing signatures of a large amount of the surrounding material;][]{katsuda2015ApJ...808...49K}.
Fe K$\alpha$ emission provides a useful diagnostic of this diversity: its centroid energy traces the ionization state of iron, while its luminosity depends on the amount and density of shocked iron.
The Fe K$\alpha$ centroid--luminosity plane of \citet{yamaguchi2014discriminating} separates Type Ia and core-collapse (CC) SNRs in a broad sense, but also shows that remnants within the same broad progenitor class span a range of Fe ionization states and Fe K$\alpha$ luminosities.
Previous 1D hydrodynamical, non-equilibrium ionization (NEI) and thermal X-ray studies have clarified how the X-ray emission of Type Ia SNRs depends on the explosion model and ambient density \citep{badenes2003thermal, badenes2005thermal, badenes2006constraints,badenes2007are, martinez2018chandrasekhar, Court_2024, court2026classification}.
These studies connected SN Ia explosion models to observed Type Ia SNR properties.
Our previous fixed-density uniform-ISM calculations of several 3D Type Ia explosion models showed that explosion-model differences can contribute to the diversity of X-ray properties; 
however, they also suggested that a uniform environment does not naturally produce the large-scale ring-like or
bipolar morphologies, or the unusually bright and highly ionized Fe K$\alpha$ properties, seen in some strongly interacting Type Ia and Type Ia-candidate remnants \citep{fujimaru2026probing}.

This spectroscopic diversity is accompanied by a wide range of morphologies.
Several Type Ia or Type Ia-candidate SNRs show pronounced asymmetries, ring-like or clumpy substructures, and evidence for interaction with dense or non-uniform CSM.
For example, N103B exhibits a double-ring structure in the intermediate-mass-element (IME) ejecta, interpreted as expansion inside an hourglass-shaped cavity created by the progenitor system \citep{yamaguchi2018evidence}.
SNR 0519--69.0 has been suggested to interact with dense circumstellar material \citep{guest2023rapid,schindelheim2024snr}, and 3C 397 shows an elongated morphology and a dense environment while its X-ray abundances support an energetic near-$M_{\rm Ch}$ Type Ia origin \citep{safi2005chandra, martinez2020evidence, treyturik2026revisiting}.
These objects suggest that the diversity of Type Ia SNRs reflects not only the explosion ejecta, but also the structured environments into which the remnants expand.

W49B is an extreme example of a SNR whose nature remains difficult to explain.
It has a centrally concentrated, bar-like metal-rich morphology, strong Fe~K emission, and Fe radiative recombination continuum, and has long been discussed as a possible jet-driven CC remnant.
However, its abundance ratios and/or metal distribution have also been argued to be consistent with an asymmetric Type Ia explosion \citep{zhou2018asymmetric, sato2025missing, treyturik2026revisiting}.
The Fe radiative recombination continuum indicates the presence of recombining, overionized plasma in W49B.
Such overionized plasma is usually discussed in mixed-morphology or CC SNRs, where dense winds or CSM first promote rapid ionization and later cooling occurs through breakout, adiabatic expansion, thermal conduction, or mixing \citep[e.g.,][]{itoh1989effect,miceli2010spatial,lopez2013unraveling,katsuragawa2022formation}.
W49B therefore illustrates how Type Ia-like abundance signatures, asymmetric morphology, and strong environmental or plasma signatures may coexist.

One possible way to realize a strongly structured circumstellar environment is a CE episode preceding the SN explosion.
CE evolution can produce non-spherical circumstellar structures, including equatorial density enhancements and bipolar morphologies \citep{garcia2018common, ondratschek2022bipolar, borges2026stars}.
If a Type Ia explosion occurs before this material has dispersed, the explosion
would take place inside a planetary-nebula-like CSM.
Such systems have been discussed as SNe Ia exploding inside planetary nebulae (PNe),
or ``SNIPs'' \citep{kashi2011circumbinary,tsebrenko2013type,tsebrenko2015modelling, Soker2024Supernovae,soker2026type}. 
The scenario is closely tied to the `core-degenerate' scenario \citep{kashi2011circumbinary}; here, the CE is formed by the merger of a C+O white dwarf (WD) and an asymptotic giant branch (AGB) star, leaving a WD-WD binary inside a PN that eventually merges to explode as an SN Ia.
Related thermonuclear explosions arising from WD--core mergers within a common envelope have also been studied in the context of their SN light curves \citep{kozyreva2024thermonuclear}.

Previous multidimensional hydrodynamical simulations have investigated the interaction of SNe Ia with non-spherical PN-like CSM and demonstrated the formation of asymmetric SNR morphologies \citep{tsebrenko2013type,tsebrenko2015modelling,chiotellis2020interaction}. 
In particular, \citet{chiotellis2020interaction} modeled Kepler's SNR as an SN Ia exploding inside a bipolar PN and reproduced some of its observed morphological and kinematic features.
\citet{Court_2024} studied this scenario with 1D hydrodynamic and NEI calculations and showed that the CE-to-SN delay time strongly affects the SNR dynamics and X-ray properties.
However, because PN-like CSM is intrinsically non-spherical, 1D calculations cannot address bipolar flows, ring-like ejecta substructures, viewing-angle effects, spatially localized overionized plasma, or the connection between velocity structure and X-ray line profiles.
Recent XRISM/Resolve observations of W49B and N103B further emphasize that velocity structure, temperature structure, and ionization state must be interpreted together in three dimensions \citep{xrism2025kinematic,holland2026xrism}.

In this work, we investigate the 3D evolution of Type Ia SNRs expanding inside non-spherical, PN-like CSM formed after CE evolution.
To explore the environmental origin of the many faces of Type Ia SNRs, we vary the characteristic scale and density of the CSM while keeping the same Type Ia explosion model.
Our aim is to clarify how a structured CSM affects the remnant morphology, plasma evolution, and X-ray properties.

The paper is organized as follows.
In Section~\ref{sec:methods}, we describe the hydrodynamical simulations, the Type Ia ejecta model, the PN-like CSM models, the NEI calculation, and the synthetic X-ray analysis.
In Section~\ref{sec:result}, we present the results. 
Section~\ref{sec:morphology} discusses the remnant morphology and ejecta distribution.
Section~\ref{sec:plasma} examines the shock-processing history, thermal evolution, and overionized plasma.
Section~\ref{sec:x-ray} presents the synthetic X-ray spectra, light curves, Doppler signatures, and Fe K$\alpha$ diagnostics.
Finally, in Section~\ref{sec:summary}, we summarize our conclusions.

\section{Methods} \label{sec:methods}

\subsection{Simulation code} \label{subsec:code}

To simulate the 3D evolution of the SNR, we use the same hydrodynamic code with NEI as in \citet{fujimaru2026probing}. 
The hydrodynamics is solved using the PPMLR scheme implemented in the multidimensional hydrodynamic code VH-1 \citep[e.g.,][]{blondin2001rayleigh}. 
The calculation is performed on a comoving Eulerian grid that expands with the SNR, making it feasible to follow the long-term evolution of the remnant over a large dynamical range \citep[for the detailed description, see][]{ferrand20103d, ferrand2019supernova}.

The local plasma states are recorded using Lagrangian tracer particles \citep[e.g.,][and references therein]{Court_2024, court2026classification}. 
For each tracer particle, the electron temperature and density histories are used to compute the NEI evolution of the shocked plasma, using ionization and recombination rate coefficients from AtomDB/APED \citep[e.g.,][]{smith2005astrophysical}. 
Each tracer particle is also assigned a component label indicating its initial origin, such as the ejecta or one of the CSM components. 
This labeling allows us to analyze the morphology, plasma state, and X-ray emission separately for each component.
Further details of the numerical method and its validation are given in \citet{fujimaru2026probing}.

Synthetic X-ray spectra are calculated using the SOXS software package \citep{zuhone2023soxs}. 
For each tracer particle, SOXS computes the thermal X-ray spectrum from the electron temperature, ionization states, density, and ion abundances, using emissivities from AtomDB/APEC \citep[e.g.,][]{smith2005astrophysical}. 
The thermal line broadening is included in this step. 
When Doppler effects are considered, the spectra generated for individual tracer particles are subsequently shifted according to their line-of-sight velocities using a separate post-processing routine. 
The spectra of selected tracer particles are then summed to obtain spectra for the whole remnant, for individual components, or for specific projected regions.

In addition, for calculations that require many epochs, such as X-ray light curves and Fe K$\alpha$ line diagnostics, we use another implementation that interpolates the AtomDB/APEC emissivity tables. 
In this implementation, emissivities tabulated as a function of electron temperature are linearly interpolated to the temperature of each tracer particle, and then weighted by the tracer-particle emission measure, elemental abundance, and ion fraction.
This implementation is optimized for rapidly computing emissivities from the plasma properties recorded by the tracer particles.

\subsection{Explosion model} \label{subsec:model_exp}

We adopt the Type Ia SN explosion model \texttt{N100ddt} from \citet{seitenzahl2013three}.
The \texttt{N100ddt} model was one of the explosion models investigated in \citet{ferrand2019supernova,ferrand2021supernova} and \citet{fujimaru2026probing}, where its evolution was followed in a uniform-ISM medium. 
In the present study, we use the same ejecta model for all simulations to isolate the effects of the PN-like CSM.

\texttt{N100ddt} is a near-Chandrasekhar-mass delayed-detonation model with an ejecta mass of $1.4\,M_\odot$, $^{56}$Ni mass of $6.04\times10^{-1}\,M_\odot$ ($\sim43\%$ of ejecta), and a kinetic energy of $1.43\times10^{51}\ {\rm erg}$. 
The model has a relatively spherical ejecta structure because it assumes a large number of ignition points, which leads to a strong deflagration phase. 
The strong deflagration causes substantial pre-expansion of the ejecta, lowering the density when the detonation front propagates. 
As a result, the model yields a relatively large amount of intermediate-mass elements and a relatively small amount of $^{56}$Ni, compared to the models having a small number of ignition points. The predicted properties of \texttt{N100ddt} agree well with those of typical SNe Ia \citep[e.g.,][]{kwok2026ApJ..1005...52K}.

The relatively spherical structure of the ejecta is useful for distinguishing the effects of the asymmetric CSM from those of intrinsic explosion asymmetry. 
For further details of the original \texttt{N100ddt} explosion model, we refer the reader to \citet{seitenzahl2013three}; the ejecta density and abundance profiles used in the SNR calculations are described in \citet{fujimaru2026probing}.

We note that the CE + core-merger scenario mentioned above in connection with SNIPs is more naturally associated with a DD/sub-$M_{\rm Ch}$ channel than with the near-$M_{\rm Ch}$ \texttt{N100ddt} model. 
However, the DD models \texttt{OneExp} and \texttt{TwoExp} \citep{pakmor2022fate} developed more asymmetric SNR morphologies in \citet{ferrand2025role} and \citet{fujimaru2026probing}, whereas \texttt{N100ddt} remained relatively spherical. 
We therefore adopt \texttt{N100ddt} here to isolate the effects of the asymmetric CSM, and leave more self-consistent progenitor--explosion combinations for future work.

\subsection{CSM models} \label{subsec:model_csm}

\begin{figure}[ht!]
\centering
\includegraphics[width=0.47\textwidth]{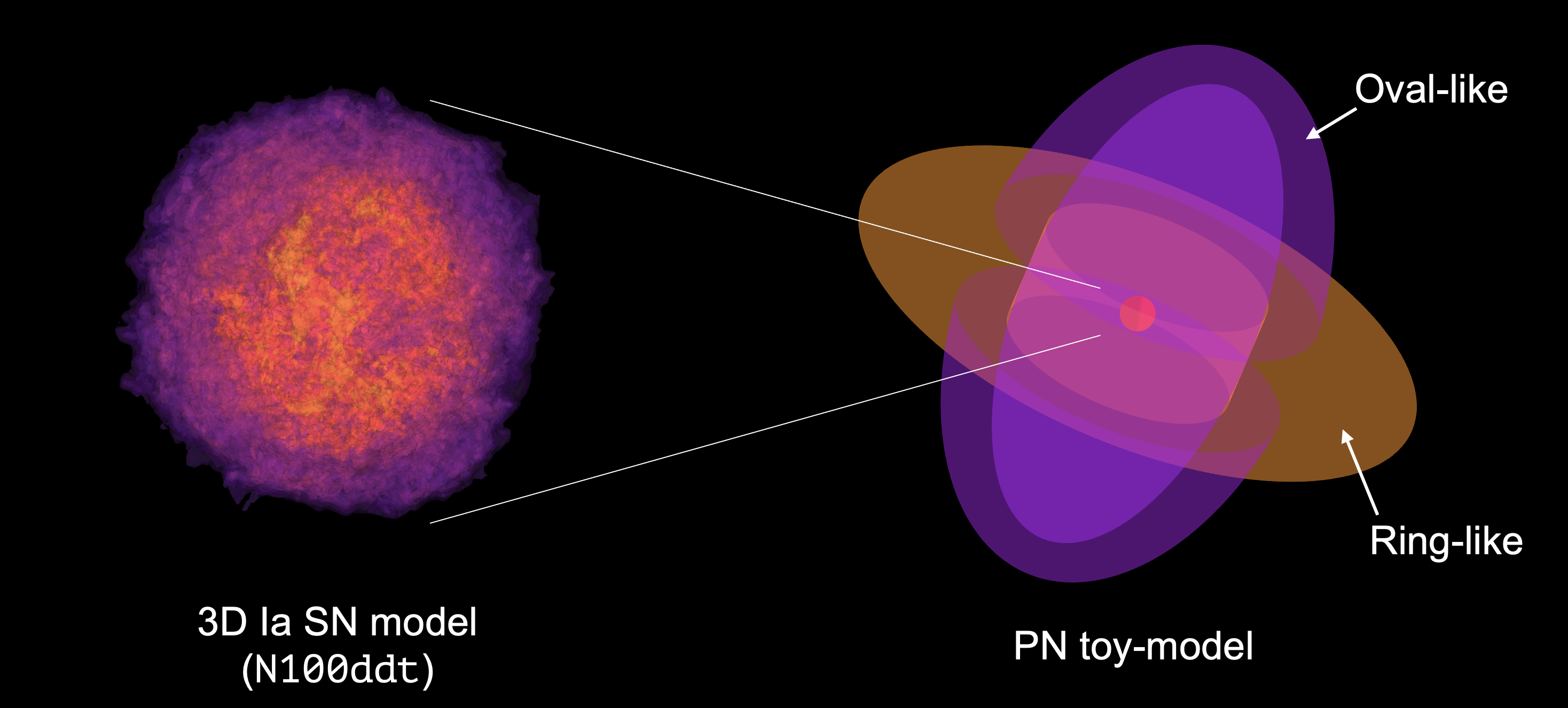}
\caption{Model setup for a Type Ia SN inside a PN. The left panel shows the freely expanding ejecta density isosurface of the \texttt{N100ddt} model \citep{seitenzahl2013three} at $t=3$ yr, when 
the maximum density is $\sim 1.5\times 10^{-18}\,\mathrm{g\,cm^{-3}}$ and the ejecta radius is $0.1\,\mathrm{pc}$. 
The right panel shows our toy PN model, based on the structure of the 2D PN simulation of \citet{garcia2018common}. 
The characteristic size scales and masses are summarized in Table~\ref{tab:modelparameter}.
\label{fig:model}
}
\end{figure}

\begin{deluxetable*}{ccccccc}
\tablewidth{0pt}
\tablecaption{CSM parameters \label{tab:modelparameter}}
\tablehead{
\colhead{\textbf{Model}} & Delay time (yr) & \colhead{$R_\text{ring}$ (pc)} & \colhead{$M_\text{ring}$ ($M_\odot$)} & \colhead{$\rho_\text{ring}$ ($\text{g} \, \mathrm{cm^{-3}}$)} & \colhead{$R_\text{oval}$ (pc)} %& \colhead{$M_\text{oval}$ ($M_\odot$)}
}
\startdata
 short & 1200 & 0.13 -- 0.3 & 1 & $5.0 \times 10^{-21}$ & 0.5 -- 0.6 & \\
 middle & 5000 & 0.5 -- 1.3 & 1 & $5.0 \times 10^{-23}$ & 1.3 -- 1.6 &  \\
 long & 10000 & 1.0 -- 2.5 & 1 & $6.7 \times 10^{-24}$ & 2.2 -- 2.5 & \\
\enddata
\tablecomments{Model parameters for the three PN delay-time models. 
$R_\mathrm{ring}$ represents the inner and outer radius of the ring-like structure, $M_{\text{ring}}$ represents the ring mass, and $R_\mathrm{oval}$ represents the inner and outer radius of the oval-like structure. 
More detailed discussions on each model are provided in the corresponding references.}
\end{deluxetable*}

We consider a Type Ia SN exploding inside a PN formed after a CE phase. 
In this scenario, the CE interaction ejects the envelope of the companion and produces a non-spherical PN-like CSM. 
After the CE phase, the remaining CO WD is assumed to grow toward the Chandrasekhar mass through subsequent accretion and eventually explodes as a Type Ia SN. 
Hereafter, the CE-to-SN delay time denotes the time interval between the formation of the PN-like CSM after the CE phase and the SN explosion, rather than the delay time from star formation to the SN.
This timescale is closely related to the common-envelope-to-explosion delay (CEED) time discussed by \citet{soker2019common,soker2022common}.

Our CSM setup is motivated by the 1D SNR-in-PN calculations of \citet{Court_2024}, 
which studied Type Ia SNRs evolving inside PN and showed that the delay time between CE ejection and the SN explosion strongly affects the SNR dynamics and X-ray properties.
Their calculations used radial density profiles taken from the post-CE PN simulations of \citet{garcia2018common}.
In the present work, we extend this scenario to a 3D, non-spherical CSM.
To construct the CSM geometry, we are guided by the post-CE PN simulations of \citet{garcia2018common}.
These simulations follow the evolution from CE ejecta to PN-like nebulae after the dynamical CE ejection phase, in which envelope material can be expelled preferentially near the orbital plane \citep{ricker2012amr}.
They explored different stellar and binary parameters and showed that post-CE material can develop strongly non-spherical morphologies.
Among their models, we adopt model A2 as a reference for the characteristic morphology and spatial scale of the CSM.
This model corresponds to a PN formed after CE mass ejection from a $2.5\,M_\odot$ AGB progenitor, and develops a dense equatorial structure with more extended lower-density regions.
We adopt the snapshots at 1200, 5000, and 10,000 yr after the CE phase as representative cases, which we refer to as the short-, middle-, and long-delay models, respectively.

We do not directly map the full hydrodynamical PN structure of \citet{garcia2018common}; instead, we construct a simplified toy model that captures its characteristic non-spherical morphology. 
The model consists of a dense equatorial ring-like component and a more extended, lower-density component surrounding the polar directions, which we refer to as the oval component. 
The remaining volume is filled with uniform ambient medium, divided into inner and outer regions relative to the ring and oval CSM structures for the purpose of component tracking. See Figure~\ref{fig:model} for a sketch of the model setup.

For each delay time, the equatorial ring mass is fixed to $M_{\rm ring}=1\,M_\odot$. 
This choice reflects the assumption that the dense equatorial CSM is mainly composed of material ejected during the CE phase, whose total mass is not expected to vary strongly with the time elapsed after ejection. 
Changing the delay time therefore primarily changes the characteristic radius and density of the CSM, with the characteristic radii based on the corresponding snapshots of \citet{garcia2018common}, rather than its total mass.
The radii of the ring and oval components are based on the corresponding snapshots of model A2 in \citet{garcia2018common}. 
In their simulations, the initially bipolar morphology can evolve toward more elliptical or barrel-like shapes through stellar-wind interactions and photoevaporation of dense equatorial material.

For simplicity, each CSM component is assigned a uniform density and is treated statically at the adopted PN age. 
The density normalization is chosen so that the equatorial ring contains $1\,M_\odot$ for each delay-time model. 
The adopted parameters of the CSM models are summarized in Table~\ref{tab:modelparameter}.
The density of the oval-like component is $0.1 \,\mathrm{cm}^{-3}$ in all models, and the corresponding masses in the short-, middle- and long-delay models are $\sim 3.0\times 10^{-4}M_\odot$, $1.0\times 10^{-2} M_\odot$, $3.4 \times 10^{-2} M_\odot$, respectively.
While these values are not tuned to accurately match the results of the PN simulations, they are intended to qualitatively represent a PN with a dense equatorial ring and a more diffuse surrounding component.
We note that the oval density would not affect our results as long as its mass is much lower than those of the other components.

\subsection{Simulation} \label{subsec:sim}

Using the code described in Section~\ref{subsec:code} and the models described in Section~\ref{subsec:model_exp} and Section~\ref{subsec:model_csm}, we calculate the hydrodynamical evolution, the NEI evolution of 30 elements, and the synthetic X-ray emission. 
%of Type Ia SNRs expanding inside PN-like CSM. 
The \texttt{N100ddt} ejecta are first allowed to expand freely until $t=3$ yr. 
At this early stage, the swept-up ambient mass is much smaller than the ejecta mass, so the ejecta evolution can be well approximated by free expansion. 
The freely expanding ejecta are then placed at the center of each PN-like CSM model, and the subsequent interaction between the ejecta and the CSM is followed up to $t=5000$ yr.

For comparison, we also use the corresponding N100ddt model expanding into a uniform ambient medium from \citet{fujimaru2026probing}.
The ambient number density of this comparison model is $n_{\rm ISM}=0.3\,{\rm cm^{-3}}$, and the model was followed up to $t=1000$ yr.
Hereafter, we refer to this model as the uniform-ISM model.

The comoving grid resolution is $256^3$. 
For the PN-like CSM simulations, we use 270,000 Lagrangian tracer particles in total: 210,000 for the ejecta, 10,000 for the equatorial ring, 10,000 for the oval component, and 10,000 and 30,000 for the inner and outer ambient media, respectively. 
Each tracer particle is initialized with a component flag, which is used to separate the ejecta and the different CSM components in the subsequent analysis.

We set the post-shock electron-proton temperature ratio, $\beta=T_e/T_p$, to 0.02. It is much larger than the mass-proportional value, $m_e/m_p$, representing efficient collisionless electron heating at the shock. 
The adopted value is within the observationally inferred range; it is somewhat smaller than the electron-proton temperature ratios inferred for some SNR shocks from observations \citep[e.g.,][]{ghavamian2013electronion,raymond2023electron}, while 
it is slightly higher than $\beta \simeq0.01$ inferred for Tycho's SNR \citep{yamaguchi2013new}.
We note that the electron-heating efficiency can affect Fe K$\alpha$ emission \citep{badenes2005thermal}. 
However, its impact is expected to be reduced in the relatively dense plasma considered here, where collisional ionization rapidly diminishes the effect of the initial collisionless electron heating at the shock \citep{court2026classification}. 
We also tested a wider range of $\beta$, from the mass-proportional value $m_e/m_p$ up to $\beta=0.1$, and found only minor differences in the thermal and ionization diagnostics presented in Figure~\ref{fig:Te_TCIE_Fe} and Figure~\ref{fig:Te_TCIE_dist_Fe}. These variations therefore do not affect our main conclusions. 

Radiative cooling is not included. 
To examine whether this assumption affects the dynamical evolution, we estimate the characteristic timescale over which the plasma would lose its internal energy through radiation. 
For the simulation as a whole, this effective cooling timescale ($t_{\rm cool}^{\rm global}$) can be evaluated as the ratio of the total internal energy to the total radiative energy-loss rate:
\begin{equation}
t_{\rm cool}^{\rm global} = \frac{\displaystyle\sum_i E_{\mathrm{int},i}}
{\displaystyle\sum_i \dot{E}_{\mathrm{cool},i}},
\end{equation}
where $E_{\mathrm{int},i}$ and $\dot{E}_{\mathrm{cool},i}$ are the internal energy and estimated radiative cooling rate of the tracer particle $i$, respectively. The resulting effective cooling timescale remained longer than the remnant age at all simulated epochs.

Because this volume-integrated estimate may conceal shorter cooling timescales in locally dense gas, we also estimate the local cooling timescale under representative conditions in the shocked ring, where radiative cooling is expected to be most effective.
The local radiative cooling timescale is written as
\begin{equation}
\label{eq:t_cool_general}
 t_\mathrm{cool} = \frac{\frac{1}{\gamma -1}n_\mathrm{tot}k_\mathrm{B}T}{n_\mathrm{e} n_\mathrm{ion}\Lambda(T)},
\end{equation}
where $\gamma$ is the adiabatic index, $n_\mathrm{tot}$ is the total particle number density, including electrons and all ion species, $n_\mathrm{e}$ and $n_\mathrm{ion}$ are the electron and total ion number densities, respectively, $k_\mathrm{B}$ is the Boltzmann constant, and $\Lambda(T)$ is the radiative cooling function, defined such that the volumetric radiative cooling rate is $n_\mathrm{e}n_\mathrm{ion}\Lambda(T)$. 
Here, we assume a single-temperature plasma, $T=T_\mathrm{e}=T_\mathrm{ion}$, and evaluate the cooling function as $T=T_\mathrm{e}$. 
Because the ions are generally hotter than the electrons, the resulting cooling timescale represents a conservative lower limit.
For a fully ionized plasma with solar abundances, Equation~(\ref{eq:t_cool_general}) can be approximated as
 \begin{equation}
\label{eq:t_cool_csm}
\begin{aligned}
 t_\mathrm{cool} 
&\simeq 3.1\frac{k_\mathrm{B}T}{n_\mathrm{e}\Lambda(T)}\\
&\simeq 1.3 \times 10^5 \, \mathrm{yr}\\
&\qquad\times\left(\frac{T}{10^6 \, \mathrm{K}}\right) \left(\frac{n_\mathrm{e}}{1 \mathrm{cm^{-3}}}\right)^{-1}\left(\frac{\Lambda(T)}{10^{-22}\, \mathrm{erg\,cm^3\,s^{-1}}}\right)^{-1} \ .
\end{aligned}
\end{equation}

Although the cooling function depends on the elemental abundances and ionization state of the plasma, adopting $T=10^{6}\,\mathrm{K}$, $n_{\mathrm e}=1\,\mathrm{cm^{-3}}$, and $\Lambda(T)=10^{-22}\,\mathrm{erg\, cm^{3}\,s^{-1}}$ as representative values for the shocked ring, where radiative cooling is expected to be most efficient, gives $t_{\mathrm{cool}}\simeq1.3\times10^{5}\,\mathrm{yr}$. 
This timescale is substantially longer than the maximum simulated age of 5000 yr. 
This order-of-magnitude estimate is also consistent with our a posteriori calculation of the volume-integrated cooling timescale, which remains longer than the remnant age throughout the simulations. 
Radiative losses are therefore not expected to significantly affect the simulated evolution, which is governed primarily by shock heating and adiabatic expansion.

For the spectral calculations, we use \texttt{SOXS} version 4.6.0 and \texttt{AtomDB} version 3.0.9. 
The spectra are calculated using 9000 linear energy bins over the 0.3--10.0 keV band.  
Unless otherwise stated, we adopt a fiducial distance of 11.3 kpc and an absorbing column density of $N_{\rm H}=5.0\times10^{22}\ {\rm cm^{-2}}$. 
These choices are motivated by the values commonly adopted for W49B \citep[e.g.,][]{zhou2018asymmetric,keohane2007near}, and are used only to generate synthetic X-ray observables. 
They do not affect the hydrodynamical or NEI evolution.
We neglect internal absorption and scattering by the CSM and ejecta in the synthetic X-ray spectra. 
To estimate the possible impact of absorption, we calculate the CSM and Fe column densities. 
At $t=3$ yr, before the equatorial ring is shocked ($t\simeq4$ yr), the unshocked CSM remains dense. 
In the short-delay model, the ring has a density of $n_{\rm H}\sim3\times10^3\,\mathrm{cm^{-3}}$ and a thickness of $\sim0.17\,\mathrm{pc}$, yielding $N_{\rm H}\sim1.6\times10^{21}\,\mathrm{cm^{-2}}$. 
Integrating the Fe mass density along lines of sight gives maximum Fe columns of $N_{\rm Fe}\sim1.2\times10^{21}$, $6.5\times10^{18}$, and $2.7\times10^{17}\,\mathrm{cm^{-2}}$ at $t=3$, 53, and 503 yr, respectively.
These estimates suggest that internal photoelectric absorption may affect early soft X-ray emission, particularly at $t=3$ yr in the short-delay model, where the dense equatorial CSM remains unshocked. 
The effect of CSM absorption is expected to be weaker in the middle- and long-delay models because of their lower CSM densities. 
Absorption by unshocked ejecta may also contribute, particularly in the Fe L band. 
Internal absorption should weaken as the remnant expands and be less important at Fe K energies. 
Preferential absorption of receding ejecta could also suppress redshifted emission, producing an apparent blueshift. 
Quantitative attenuation depends on the composition and ionization state of the absorbing material, which we do not model.

\section{Results and Discussion} \label{sec:result}

\subsection{Morphology} \label{sec:morphology}

\begin{figure}[ht!]
\includegraphics[width=\columnwidth]{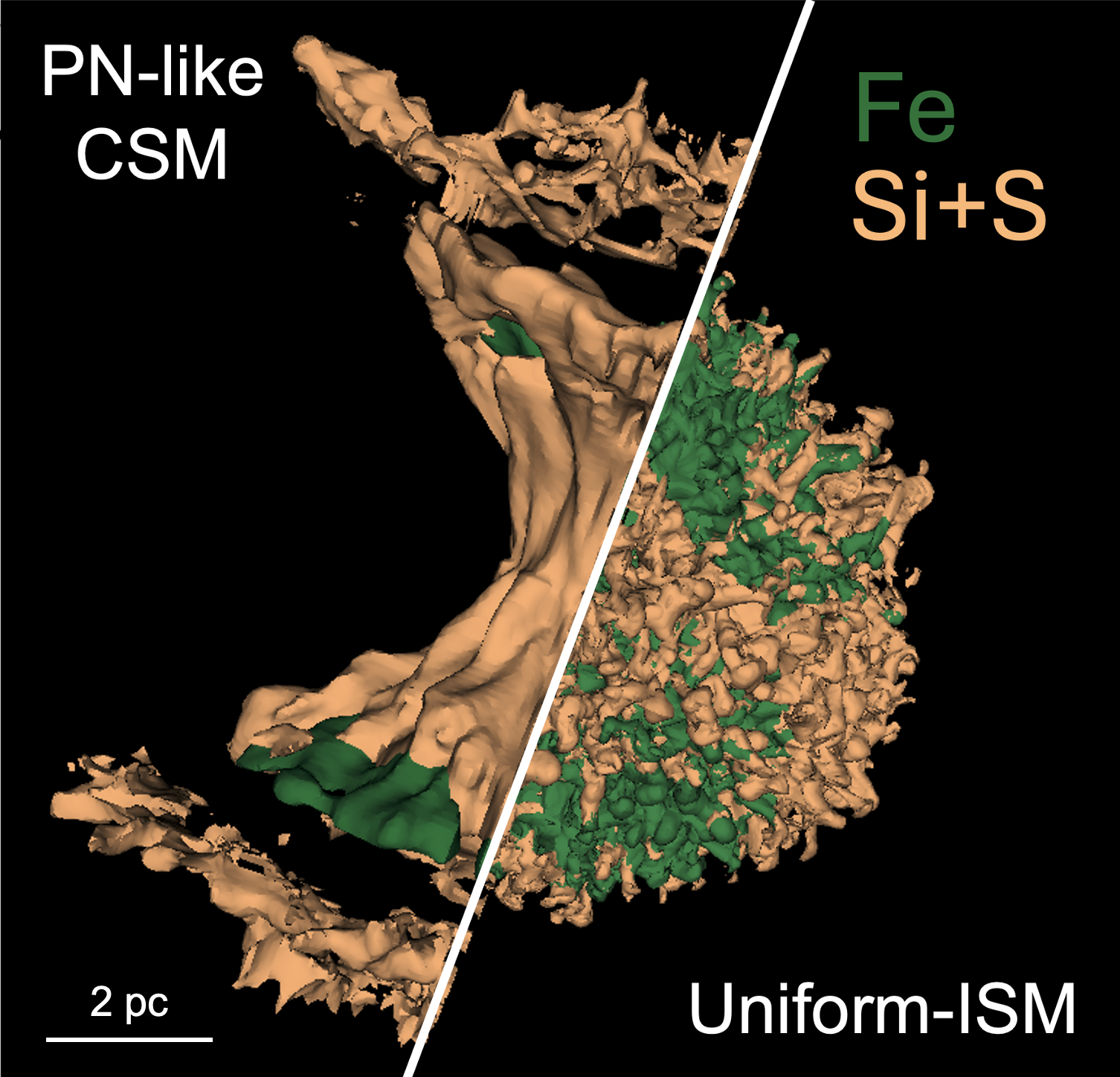}
\caption{Isosurfaces of Fe and Si+S densities. The isosurface values are set to 10\% of the maximum density.
Short-delay model at 500 yr is shown in the left side, and the uniform ambient density ($n_{\rm ISM}=0.3\,{\rm cm^{-3}}$) model at 400 yr \citep{fujimaru2026probing} is shown in the right side. 
\label{fig:isosurface_ring_and_uni}}
\end{figure}

\begin{figure*}[ht!]
\includegraphics[width=\textwidth]{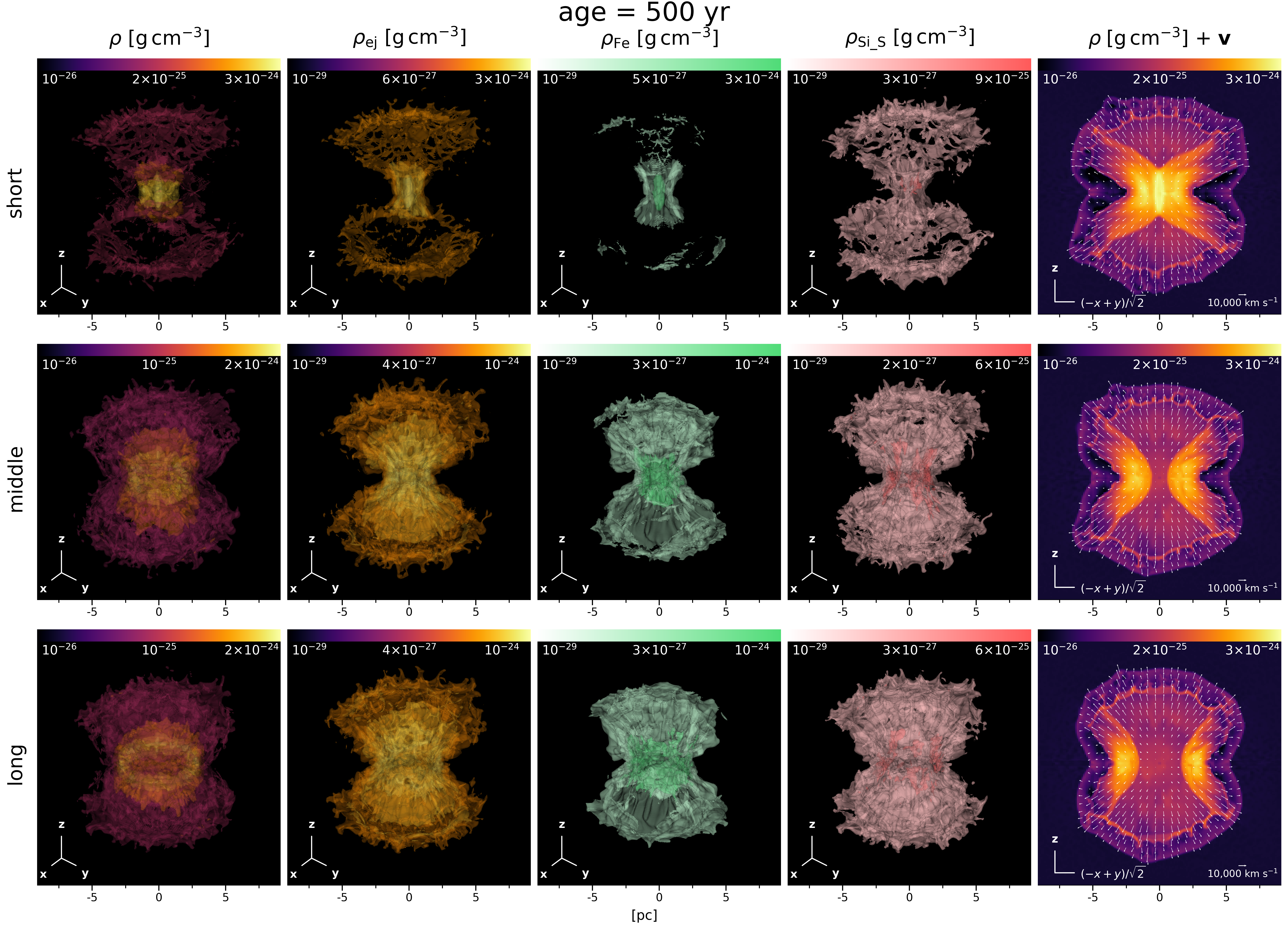}
\caption{Density isosurfaces and a density slice at $t=500\,\mathrm{yr}$. 
Each row corresponds to a PN model, and each column to a different component. 
From left to right, the panels show the total density, ejecta density, Fe density, Si+S density, and a slice of the total density overlaid with velocity vectors. 
The isosurface levels for the total density and ejecta density panels are set to 5\%, 25\%, 50\% and 75\% of their respective maximum densities, whereas those for the Fe and Si+S panels are set to 5\% and 50\% of the maximum density, respectively.
The isosurface panels are viewed from $(\theta,\phi)=(60^\circ,45^\circ)$, and the normal to the slice plane in the rightmost column is $(\theta,\phi)=(90^\circ,45^\circ)$.
Here, $\theta$ is measured from $z$-axis, and $\phi$ is the azimuthal angle measured from the positive $x$-axis in the $xy$-plane.
\label{fig:isosurface_500}}
\end{figure*}

\begin{figure*}[ht!]
\includegraphics[width=\textwidth]{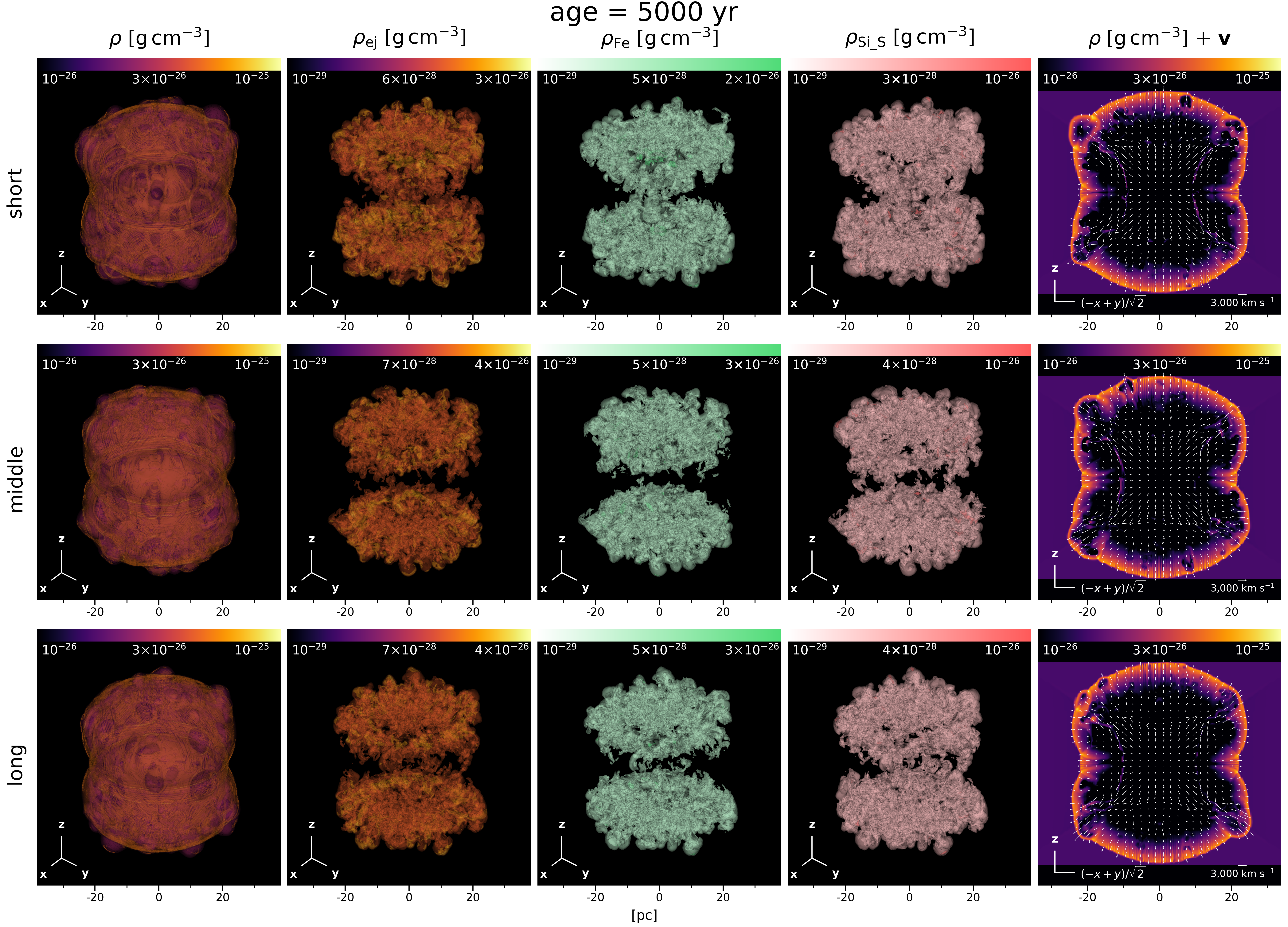}
\caption{Density isosurfaces and a density slice at $t=5000\,\mathrm{yr}$. 
The panel layout, and the isosurface definitions and viewing angles are the same as in Figure~\ref{fig:isosurface_500}.
\label{fig:isosurface_5000}}
\end{figure*}

We first examine whether our scenario can naturally produce morphologically diverse remnants from the same explosion model. 
In particular, we focus on how changing the characteristic PN scale, parameterized by the delay time after the CE phase, affects the global remnant morphology and the ejecta/elemental distributions. 
Because the same explosion model is used for all simulations, differences among the three models reflect differences in the CSM scale and density. 
The relatively spherical structure of the \texttt{N100ddt} ejecta further helps us identify large-scale deformation produced by the CSM interaction.

Figure~\ref{fig:isosurface_ring_and_uni} illustrates the effect of the PN-like CSM by comparing the short-delay model with a model evolving in a uniform-ISM medium at a comparable young age. 
In the uniform case, the ejecta still show asymmetries caused by the intrinsic abundance structure of the explosion and by the growth of hydrodynamical instabilities. 
However, these structures are much weaker than the large-scale deformation seen in the short-delay PN model. 
In the short-delay model, the dense and compact equatorial ring-like structure suppresses the expansion in the equatorial direction and redirects the flow toward the polar directions, producing a pronounced bipolar morphology and ring-like ejecta structures associated with the bipolar flow. 
This comparison shows that a CSM with a mass of order $1\,M_\odot$ can substantially reshape the ejecta distribution during the young SNR phase.

Figure~\ref{fig:isosurface_500} and Figure~\ref{fig:isosurface_5000} show the density isosurfaces and density slices overlaid with in-plane velocity vectors for the three PN models at $t=500$ and $5000$ yr, respectively. 
At $t=500$ yr, all models have already interacted with the ring-like CSM, and the imprint of this interaction is visible in the total density, ejecta density, Fe distribution, and Si+S distribution. 
The strength of the deformation decreases from the short- to the long-delay model. 
In the short-delay model, the ring is compact and dense, so the ejecta collide with it at an early stage and the equatorial expansion is strongly suppressed. 
The density slices also show local density enhancements caused by compression of the shocked material near the equatorial ring.
As shown by the velocity vectors in the rightmost panel of Figure~\ref{fig:isosurface_500}, the flow is also deflected around the dense equatorial ring and redirected toward the polar directions, producing a pronounced bipolar structure.
In the middle- and long-delay models, the ring is located at larger radii and has a lower density, so the interaction occurs later and produces a weaker distortion.
At $t=5000$ yr, the velocity vectors in Figure~\ref{fig:isosurface_5000} more clearly reveal the collimated bipolar flow. 
In addition, the flow along the bipolar axis shows substantial deceleration associated with reflected shocks generated by the interaction of the bipolar outflow with the outer shocked structure.

The bipolar interaction also produces ring-like structures around the polar directions. 
At $t=500$ yr, these structures are already visible, especially in the short-delay model, although their appearance differs among the density components. 
By $t=5000$ yr, large ring-like structures around the upper and lower bipolar lobes are visible in all models. 
The structure is formed by the redirected bipolar flow that interacts with swept-up material near the polar lobes. 
The interaction deflects part of the flow around the lobe, producing a toroidal-like structure that appears ring-like in projection.

Similar hydrodynamic behavior has been reported in previous simulations of Type Ia SNRs interacting with PN-like CSM. \citet{tsebrenko2015modelling} showed that interactions with a PN shell and embedded clumps cause substantial shock deceleration and non-radial flows around the clumps. \citet{chiotellis2020interaction} demonstrated that shock breakout from asymmetric, PN-shaped CSM can produce equatorial ear-like protrusions, as observed in Kepler's SNR. Our simulations similarly show flow deceleration, compression, and deflection around dense CSM structures, although the resulting large-scale morphologies differ depending on the adopted CSM geometry.

By $t=5000$ yr, the morphological differences among the three models become less pronounced. 
The remnants have expanded beyond the initial dense CSM structures and have swept up more ambient material, so the large-scale shapes become less sensitive to the initial differences in the ring radius and density.
The memory of the early interaction is not completely erased, but it appears mainly as modest differences in the detailed shape and density distribution.

The elemental distributions reflect the radial stratification of the ejecta. 
Fe, which is initially concentrated in the inner ejecta, remains more centrally located at early times, although it is already distorted in the short-delay model. 
By contrast, Si and S are distributed farther out and therefore respond more directly to the early interaction with the CSM. 
This produces more extended and asymmetric Si+S structures, especially in the short-delay model. 
The element-dependent appearance of the ring-like structures therefore reflects the different shock-processing times of the Fe-rich and IME-rich ejecta layers.

These results show that the interaction of Type Ia ejecta with a PN-like CSM can produce a wide range of remnant morphologies, including bipolar and ring-like ejecta structures.
A short delay time leads to a compact and dense CSM, which strongly redirects the ejecta and produces a pronounced bipolar remnant. 
A longer delay time leads to a larger and lower-density CSM, so the same explosion model evolves into a remnant with weaker early-time deformation. 
Thus, the morphological diversity does not necessarily require different explosion mechanisms; it can arise from differences in the CSM scale set by the time interval between PN formation and the SN explosion.

Observed Type Ia and Type Ia-candidate SNRs also show a wide range of asymmetric morphologies, from ring-like structures to strongly bipolar or elongated remnants and more moderate CSM-shaped asymmetries. 
Examples include N103B for ring-like structures, W49B and 3C 397 for stronger bipolar or elongated asymmetries, and Kepler's SNR for a more moderate but still non-uniform CSM-shaped morphology \citep[e.g.,][]{yamaguchi2021discovery,zhou2018asymmetric, xrism2025kinematic, martinez2020evidence}.
SNR 0519--69.0 also shows evidence for interaction with dense CSM, for which a PN-like cocoon has been proposed \citep{guest2023rapid,schindelheim2024snr}.
Although our fiducial toy models are not intended for direct comparisons, they show that interaction with a PN-like CSM can produce a similar range of morphological ``faces'', including bipolar and ring-like ejecta structures, even from a relatively spherical Type Ia ejecta model.

\subsection{Shock processing and Plasma properties}
\label{sec:plasma}

\subsubsection{Plasma evolution}

\begin{figure*}[ht!]
\includegraphics[width=\textwidth]{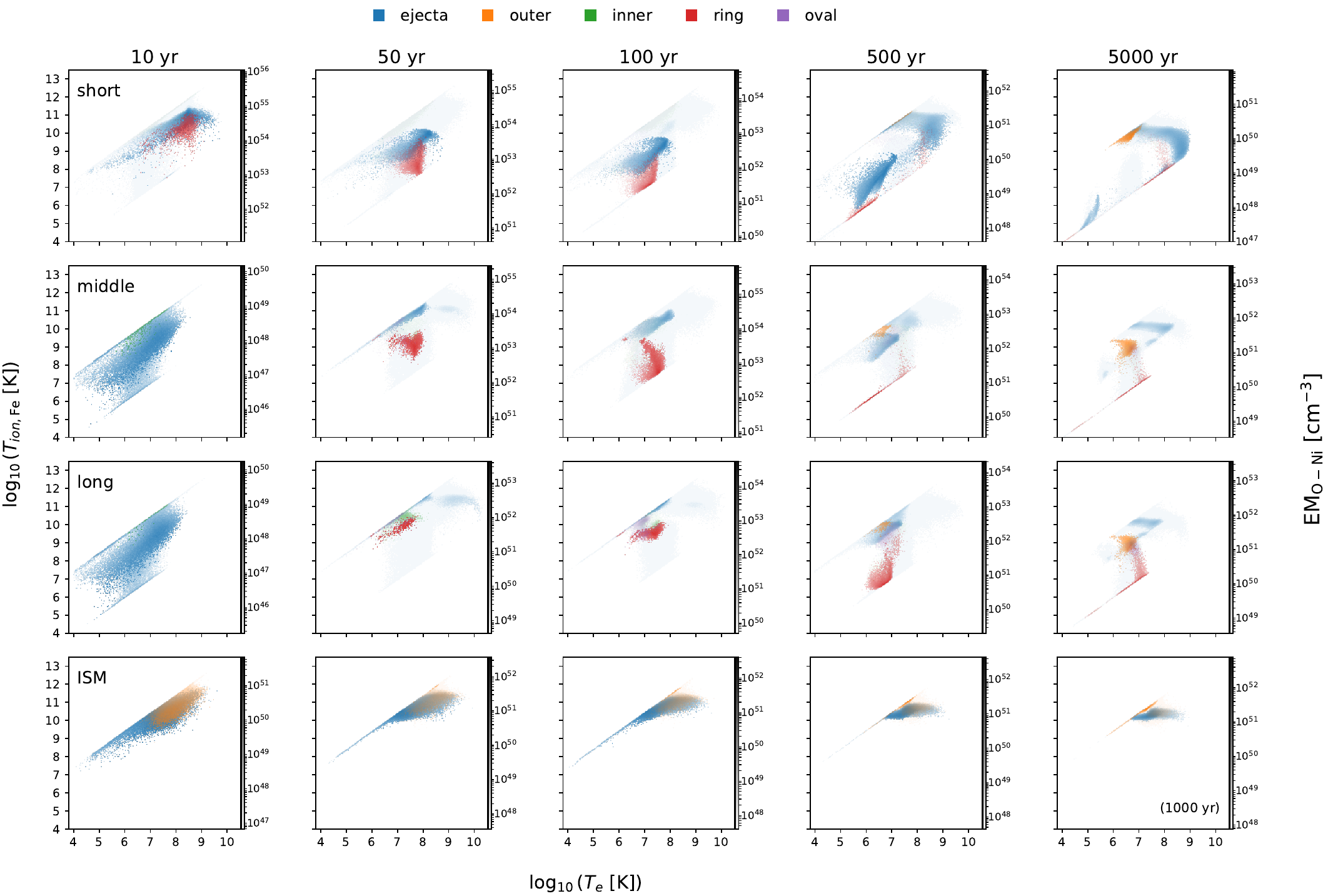}
\caption{Emission measure distributions in the $T_e$--$T_{\rm ion,Fe}$ plane. 
Here $T_e$ is the electron temperature and $T_{\rm ion,Fe}$ is the ion temperature of Fe. 
Rows correspond to the short-, middle-, and long-delay models from top to bottom, and columns correspond to different epochs, from left to right: $t=10, 50, 100, 500$ and $5000$ yr. 
The bottom row shows the uniform-ISM model for comparison; because this model was
followed only up to $t=1000$ yr, its $t=1000$ yr distribution is shown in the
rightmost column for reference.
Colors indicate the initial component to which each tracer particle belongs: ejecta, outer medium, inner medium, ring-like, and oval-like component.
The color opacity represents the EM per temperature bin, with a separate scale for each panel.
In the uniform-ISM model, orange regions denote uniform-ISM components.
The upper envelope of the distribution reflects freshly shocked plasma for which ions are hotter than electrons owing to the adopted post-shock electron--ion temperature partition, whereas plasma approaches the lower boundary near $T_e\simeq T_{\rm ion,Fe}$ as electron--ion thermal equilibration proceeds.
\label{fig:Tion_Te_Fe}}
\end{figure*}

Figure~\ref{fig:Tion_Te_Fe} shows the emission measure (EM) distributions of shocked plasma in the $T_{\rm e}$--$T_{\rm ion,Fe}$ plane.
The EM associated with a selected ion is defined as
\begin{equation}
\mathrm{EM}_{\rm ion} = \int n_{\rm e} n_{\rm ion}\,dV,
\label{eq:EM}
\end{equation}
where $n_{\rm e}$ and $n_{\rm ion}$ are the electron and ion number densities, respectively.
The EM in each two-dimensional temperature bin is obtained by summing the contributions from the corresponding tracer particles.
For Figure~\ref{fig:Tion_Te_Fe}, the EM is summed over elements from O to Ni and their ionization states.
The figure should be read first from left to right along each row, which shows the time evolution of a given delay-time model, and then from top to bottom, which shows how the evolution changes as the PN-like CSM has a larger characteristic radius and a lower density. 
Here $T_{\rm ion,Fe}$ is the Fe ion temperature.
This diagram provides a compact view of the post-shock thermal history of the plasma, including newly shocked particles, particles evolving through electron--ion thermal equilibration, and adiabatically cooled particles.
To examine the spatial correspondence of these thermal components within the ejecta, we refer the reader to Appendix~\ref{sec:append_spec_comp}, where their component-by-component spatial distributions are shown for the short-delay model.

In SNRs, shocks are collisionless, and electrons, protons, and ions are not instantaneously heated to the same temperature. 
If ions are heated approximately in proportion to their masses, the post-shock Fe ion temperature is much higher than the proton temperature. 
In the present calculations, the post-shock electron temperature is set by the adopted electron-to-proton temperature ratio, $\beta=T_e/T_p=0.02$, which is much larger than the mass-proportional electron-heating value $m_e/m_p$. 
For freshly shocked Fe, the expected temperature contrast is therefore approximately
\begin{equation}
\frac{T_{\rm ion,Fe}}{T_e} \sim \frac{A_{\rm Fe}}{\beta},
\end{equation}
where $A_{\rm Fe}\simeq56$ is the atomic mass number of Fe. 
This relation defines the upper boundary of the particle distribution in Figure~\ref{fig:Tion_Te_Fe}.
Additionally, the broad range of ion temperatures along this upper boundary reflects the range of shock velocities experienced by the freshly shocked ejecta.
In particular, the fastest outer ejecta initially have velocities of $\sim 4\times10^4\,\mathrm{km\,s^{-1}}$, and their interaction with the CSM can heat Fe ions to $T_{\rm ion,Fe}\gtrsim10^{12}\,\mathrm{K}$.
However, the Fe mass associated with such extremely high ion temperatures is very small. 
Considering all tracer components at five sampled epochs, the maximum Fe masses are $\sim1.2\times10^{-9}$, $6.8\times10^{-11}$, and $1.6\times10^{-10}\,M_\odot$ in the short-, middle-, and long-delay models, respectively, occurring at $t\simeq13$, 53, and 53 yr.
In the uniform-ISM model, these high-temperature components gradually disappear as the characteristic velocity of the ejecta being shocked decreases with time.
In the PN models, by contrast, the ejecta encounter the spatially extended and non-spherical CSM at different times, allowing high-velocity ejecta to be shock-heated over a longer period and thereby maintaining the high-temperature components for longer.
Such CSM interactions may also enhance the reverse-shock heating of Fe-rich ejecta compared to typical Type Ia SNRs evolving in a uniform ISM, potentially leading to broader Fe emission lines through thermal Doppler broadening. 
However, the observable contribution depends on the emission measure and ionization state of the heated Fe-rich plasma.

After shock heating, Coulomb collisions gradually drive electrons and ions toward thermal equilibrium. 
If Coulomb equilibration were the only important post-shock process, particles would move mainly toward the electron--ion equilibrium line, $T_e\simeq T_{\rm ion,Fe}$: electrons would be heated by ions, increasing $T_e$, while Fe ions would cool, decreasing $T_{\rm ion,Fe}$. 
In the $T_e$--$T_{\rm ion,Fe}$ plane, this corresponds to motion roughly toward the lower-right direction. 
As a reference, in a uniform ambient-medium model examined up to $t=1000$ yr, shocked particles mainly follow this lower-rightward trend, although they do not necessarily reach the equilibrium line within the SNR age because the Coulomb equilibration timescale is long.

The PN-like CSM models differ from this simple behavior because a significant fraction of the ejecta is processed in much denser CSM. 
The higher density shortens the Coulomb equilibration and ionization timescales, while the subsequent anisotropic expansion introduces strong adiabatic cooling. 
In particular, a substantial fraction of particles in the short-delay model extends toward lower $T_e$. 
This leftward extension cannot be explained by Coulomb equilibration alone and indicates adiabatic cooling of the shocked ejecta during rapid expansion. 
In the short-delay model, the compact and dense ring-like structure redirects the shocked ejecta toward the polar directions while the ejecta are still expanding rapidly, producing a prominent cooling branch in the $T_e$--$T_{\rm ion,Fe}$ plane.

Thus, the shocked plasma does not follow a single monotonic thermal path in the PN-like CSM. 
Instead, the CSM interaction produces multiple thermal components: freshly shocked plasma near the upper envelope, plasma evolving through electron--ion equilibration, and adiabatically cooled plasma at lower temperatures. 
The three PN models show similar thermal pathways, but the timing and relative importance of each component depend on the delay time. 
In the short-delay model, shock heating and adiabatic cooling occur at earlier epochs because the ejecta encounter the compact dense ring soon after the explosion. 
In the middle- and long-delay models, the same sequence appears later and more gradually, reflecting the larger characteristic radius and lower density of the CSM. 
Therefore, the delay time controls not only the amount of shocked ejecta, but also the post-shock thermal pathway of the plasma. 
This thermal history provides the physical basis for the overionized plasma discussed in Section~\ref{sec:overionized}.

\subsubsection{Shocked mass evolution}

\begin{figure}[ht!]
\includegraphics[width=\columnwidth]{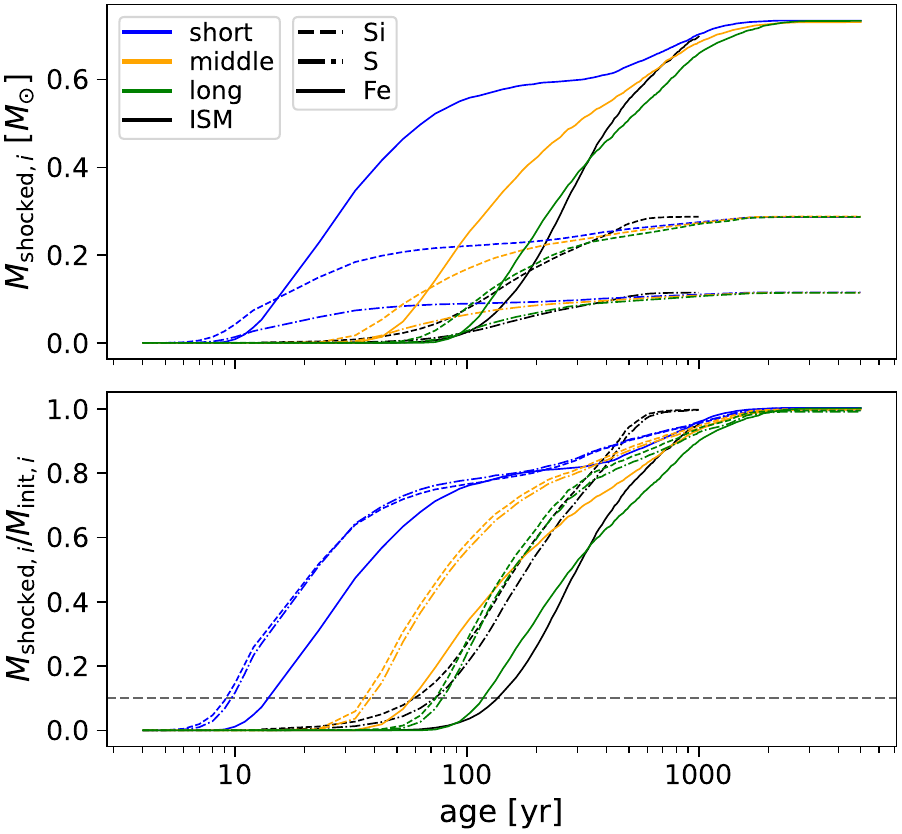}
\caption{Time evolution of the shocked masses and shocked mass fractions of Si, S, and Fe for ejecta. 
For comparison, the corresponding uniform-ISM case ($n_{\rm ISM}=0.3\,{\rm cm^{-3}}$) is also shown and labeled as ISM.
The gray dashed line marks the 10\% shocked-mass-fraction threshold used in Figure~\ref{fig:Te_TCIE_Fe}.
\label{fig:shocked_evo}}
\end{figure}

Figure~\ref{fig:shocked_evo} shows the time evolution of the shocked masses and shocked mass fractions of Si, S, and Fe. 
The key quantity is the rise time of the shocked mass fraction, which indicates when each ejecta layer is processed by the reverse shock.

The increase in the shocked mass is most rapid in the short-delay model, where the ejecta collide with the compact and dense equatorial ring at an early stage. 
The rise becomes progressively slower from the short- to the middle- and long-delay models, and is slowest in the uniform ambient-medium case. 
This ordering reflects the decreasing strength and increasing characteristic radius of the ejecta--CSM interaction: a compact dense ring drives an early reverse shock, whereas a larger, lower density ring or a uniform-ISM medium processes the ejecta more gradually.

Although the delay time changes when and how rapidly the shocked fractions increase, the ordering of the shocked elements is broadly similar among the models. 
Si and S, which are distributed farther out in the ejecta, are shocked earlier than Fe. 
This reflects the underlying radial stratification of the ejecta: the outer IME-rich layers are processed first, while the inner Fe-rich material is shocked later.

These trends show that the delay time controls not only the global morphology but also the shock-processing history of the ejecta. 
This shock history provides the basis for the different plasma and ionization states discussed below.

\subsubsection{Overionized plasma}
\label{sec:overionized}

\begin{figure*}[ht!]
\includegraphics[width=\textwidth]{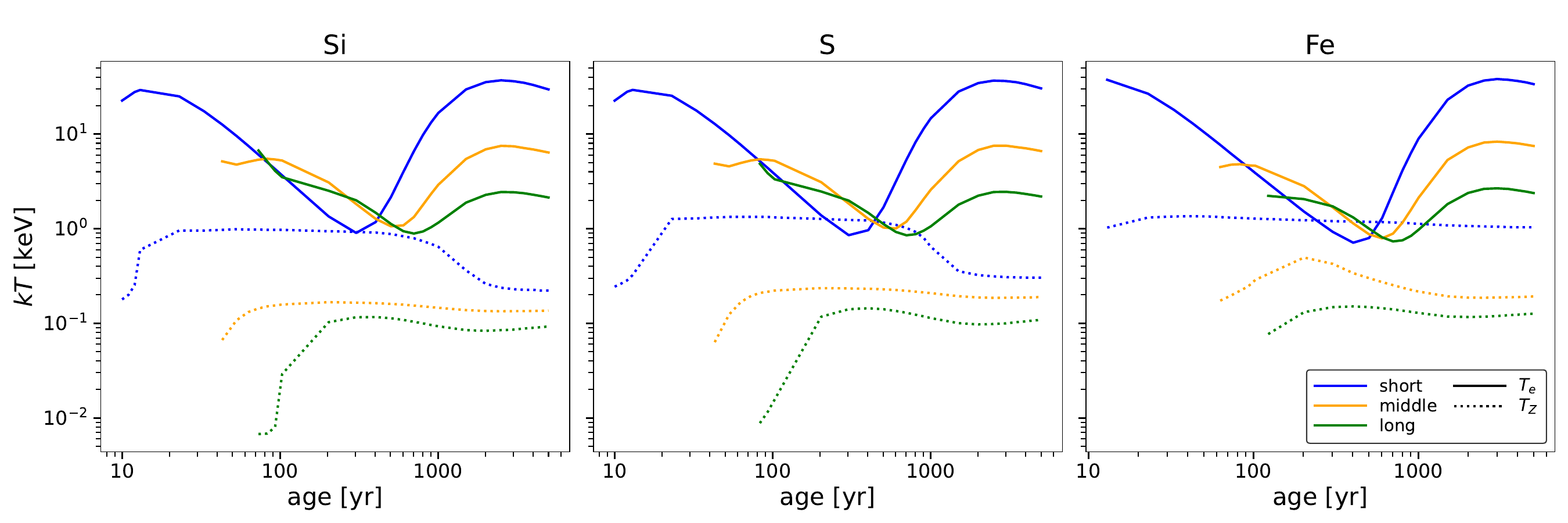}
\caption{Time evolution of the EM-weighted mean electron temperature and ionization temperature in the ejecta. 
From left to right, the panels show Si, S, and Fe, respectively. Colors indicate the three PN models. 
Solid and dotted lines denote the EM-weighted mean electron temperature and ionization temperature, respectively. The ionization temperature is defined as the equivalent CIE temperature inferred from the ionization state.
Only epochs after the shocked mass fraction of the corresponding element exceeds 10\% (Figure~\ref{fig:shocked_evo}) are shown, in order to avoid averages dominated by a very small amount of newly shocked material.
\label{fig:Te_TCIE_Fe}}
\end{figure*}

\begin{figure*}[ht!]
\includegraphics[width=\textwidth]{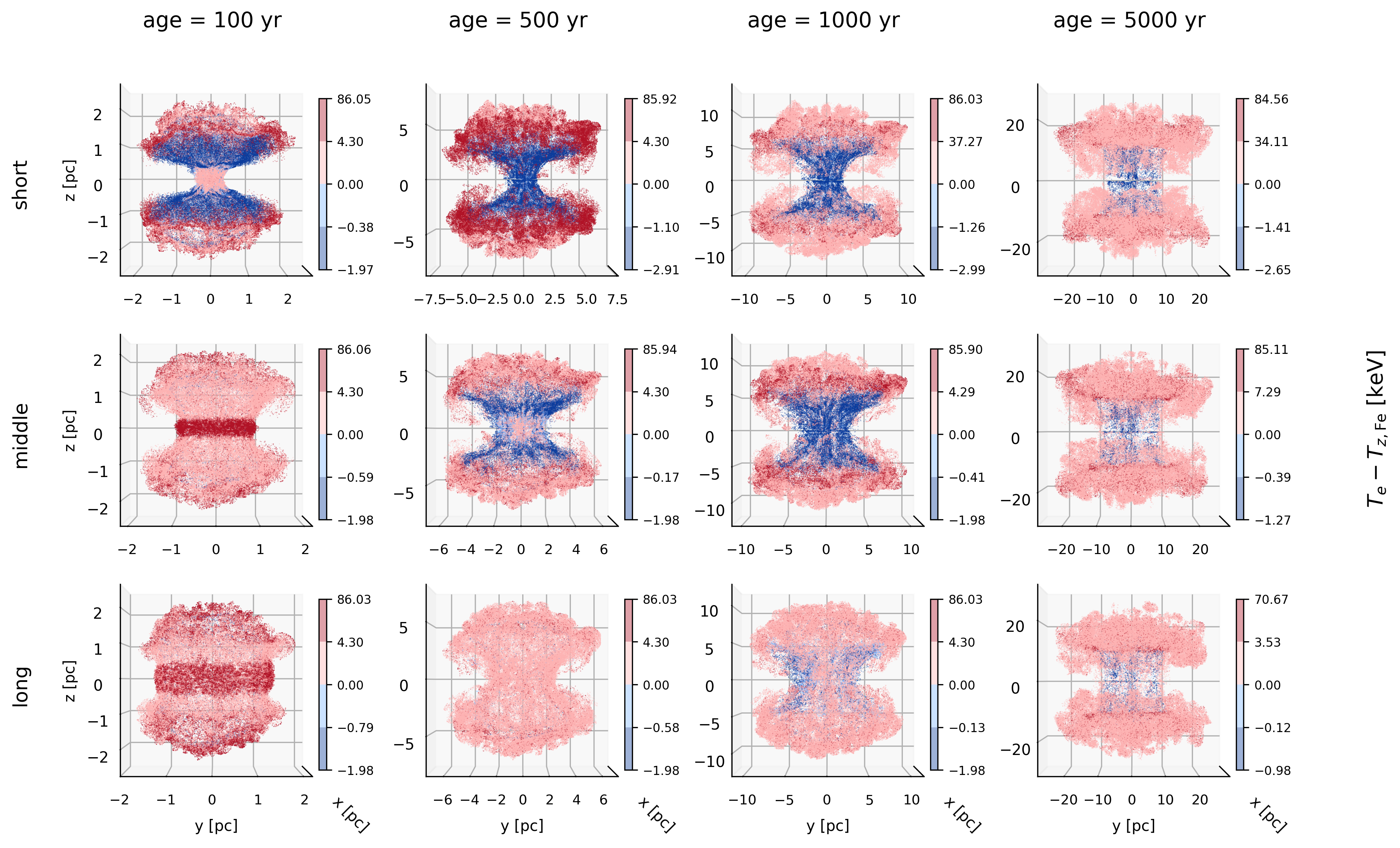}
\caption{The 3D distributions of $T_e-T_{\mathrm{Z},\mathrm{Fe}}$ for tracer particles in the ejecta. The line-of-sight is along the $x$-axis, with the observer on the positive $x$ side. Red and pink indicate $T_e > T_{\mathrm{Z},\mathrm{Fe}}$. Blue and sky blue indicate $T_e < T_{\mathrm{Z},\mathrm{Fe}}$, i.e., overionized plasma. Only tracer particles with electron temperatures within the temperature range covered by AtomDB are shown.
The excluded particles account for at most $\sim1\%$ of the total EM.
\label{fig:Te_TCIE_dist_Fe}}
\end{figure*}

To assess both the global significance and the spatial origin of overionized plasma in the shocked ejecta, we characterize the thermal and ionization states of the plasma. 
For the volume-integrated analysis, we weight the thermal and ionization properties by the EM defined in Equation~(\ref{eq:EM}). 
We use the EM of each ionization state as an indicator of its relative contribution to the thermal X-ray emission.
The EM for an element or a group of ions can similarly be obtained by summing the contributions from the relevant ionization states.
We also estimate the ionization temperature following the basic idea of \citet{katsuragawa2022formation}, but with a different implementation. 
While \citet{katsuragawa2022formation} determined the ionization temperature from the ion population pattern, we define it using the EM-weighted mean charge state. 
The ionization temperature $T_\mathrm{Z}$ is defined as the collisional ionization equilibrium (CIE) temperature at which the mean charge state is equal to the ion-EM-weighted mean charge state of the simulated plasma.
For each element $X$, this is expressed as
\begin{equation}
T_{\mathrm{Z}, X} = T_{\rm CIE}(\langle Z_X \rangle_{\rm EM_{ion}}),
\end{equation}
where $T_{\rm CIE}$ converts the ion-EM-weighted mean charge state $\langle Z_X \rangle_{\rm EM_{ion}}$ to the equivalent CIE temperature.
Using these quantities, we employ two complementary diagnostics:
the volume-integrated evolution of the electron and ionization temperatures (Figure~\ref{fig:Te_TCIE_Fe}), which indicates whether overionized plasma makes a substantial contribution to the spatially integrated X-ray emission, and the spatial distribution of the local thermal--ionization imbalance for Fe (Figure~\ref{fig:Te_TCIE_dist_Fe}), which allows us to identify where overionized plasma is produced and to clarify its physical origin.

Figure~\ref{fig:Te_TCIE_Fe} shows the time evolution of the EM-weighted mean electron temperature and ionization temperature for Si, S, and Fe in the ejecta. 
These are global diagnostic quantities and do not imply that the ejecta can be represented by a single-temperature plasma.
The key diagnostic is the relative position of the two curves: the plasma is overionized when the ionization temperature\footnote{The ionization temperature $T_Z$ should not be confused with the ion temperature $T_{\rm ion}$.} exceeds the electron temperature, i.e., when $T_\mathrm{Z} > T_\mathrm{e}$.
Only epochs after the shocked mass fraction of the corresponding element exceeds 10\% are shown, in order to avoid averages dominated by a very small amount of newly shocked material. 

In standard NEI evolution, the plasma is usually underionized relative to the CIE state expected from $T_e$, i.e., $T_e > T_\mathrm{Z}$.
If this relation is reversed, $T_e < T_\mathrm{Z}$, the plasma has a mean charge state higher than that expected for CIE at the same electron temperature, indicating an overionized state.

In all models, the EM-weighted temperature evolution can be understood as a sequence of three stages:
(1) rapid ionization and electron heating when the ejecta are shocked,
(2) subsequent electron cooling of the shocked ejecta through adiabatic expansion, and
(3) a later shift back toward the normal-NEI regime, caused by a combination of reheating by reflected shocks associated with the bipolar interaction and an increasing emission-measure contribution from normally ionizing shocked plasma.
The spatial origin of these components is shown for Fe in Figure~\ref{fig:Te_TCIE_dist_Fe}.
The timing of these stages shifts to later epochs for longer delay-time models, and these changes occur more gradually.

In the short-delay model, the first stage occurs early because the ejecta collide with the compact and dense ring-like structure. 
The high density of the shocked plasma rapidly increases $T_\mathrm{Z}$ because of efficient collisional ionization. 
At the same time, the ring suppresses equatorial expansion and redirects the shocked ejecta toward the polar directions. 
Because this occurs while the ejecta are still expanding rapidly, the subsequent adiabatic expansion efficiently lowers $T_e$, producing the phase with $T_e<T_\mathrm{Z}$.

In the middle- and long-delay models, the CSM is more extended and lower density, so the interaction occurs later and is weaker. 
The rise in $T_\mathrm{Z}$ and the decline in $T_e$ are therefore more gradual, and the EM-weighted mean plasma state does not enter a strong overionized regime.

The reversal in the short-delay model indicates that overionized plasma has a substantial influence on the global EM-weighted plasma state.
This suggests that recombining-plasma signatures can appear even in volume-integrated spectra. 
We note that the EM-weighted mean quantities can underestimate the importance of overionization, by 
hiding localized overionized regions, especially in the models where the global mean does not satisfy $T_e<T_\mathrm{Z}$. 
Therefore, the absence of a clear reversal in the middle- and long-delay models does not necessarily imply that overionized plasma is absent everywhere in the remnant.
Additionally, the global reversal in the short-delay model lasts only for a relatively short period, whereas localized overionized components can persist to ages of several thousand years in our simulations.
Whether these components produce clear signatures in the integrated spectrum depends on their emission-measure contribution relative to the other plasma components.
To reveal these spatially localized components, we examine the 3D distribution of $T_e-T_\mathrm{Z, Fe}$ in the ejecta in Figure~\ref{fig:Te_TCIE_dist_Fe}.

Figure~\ref{fig:Te_TCIE_dist_Fe} shows the 3D distribution of the same temperature imbalance discussed in Figure~\ref{fig:Te_TCIE_Fe}, using $T_e-T_{\mathrm{Z,Fe}}$ for ejecta tracer particles. 
Blue colors indicate particles with $T_e<T_{\mathrm{Z,Fe}}$, corresponding to overionized plasma, whereas red colors indicate particles with $T_e>T_{\mathrm{Z,Fe}}$.
This figure complements the EM-weighted evolution shown in Figure~\ref{fig:Te_TCIE_Fe} by revealing where the overionized and reheated plasma components are located within the remnant.

Overionized plasma is mainly found in the central and inner regions, as well as in the rapidly expanding bipolar flow. 
This spatial distribution supports the interpretation that the ejecta are first rapidly ionized when they collide with the dense ring-like CSM and are then cooled by rapid adiabatic expansion. 
As a result, the electron temperature drops below the ionization temperature expected from the ionization state, producing overionized plasma.

At larger distances along the bipolar axis, the outflow collides with the outer shocked structure and generates reflected shocks. 
These reflected shocks reheat part of the ejecta, producing regions with $T_e>T_{\mathrm{Z},\mathrm{Fe}}$ near the outer bipolar structures. 
The same bipolar flow also produces ring-like structures near each pole, which are visible in the tracer-particle distribution.

The degree of overionization and the morphology both depend on the delay time through the PN density and size scale. 
In the short-delay model, the dense and compact ring strongly collimates the ejecta, producing a pronounced bipolar morphology and strong overionization. 
In contrast, the longer-delay models experience weaker PN interaction because the CSM is more extended and lower density, resulting in less prominent overionized regions and a morphology closer to spherical.

The strong spatial variation of $T_e-T_{\mathrm{Z},\mathrm{Fe}}$ also implies that different regions of the same remnant can show distinct plasma signatures in their X-ray spectra. 
In particular, overionized features can be enhanced in regions dominated by rapidly cooled ejecta, while reheated shell regions may show hotter plasma components. 
This motivates the region-specific spectral analysis presented below, where we demonstrate this effect for the short-delay model.

These results provide a possible Type Ia route to recombining plasma produced by adiabatic cooling. 
In our models, the ejecta are first rapidly ionized through interaction with the dense equatorial CSM, and the electron temperature subsequently decreases as the shocked ejecta expand along the bipolar directions. 
This sequence naturally produces overionized plasma with $T_{\rm e}<T_Z$. 
A similar adiabatic-cooling origin has been discussed for W49B, where recombining plasma is spatially non-uniform and is enhanced toward the center and western side of the remnant \citep{miceli2010spatial, lopez2013unraveling, yamaguchi2018evidence}.

At the same time, the presence of such a thermal history does not guarantee that the plasma will remain overionized at the observed epoch. 
Some ejecta that have undergone rapid ionization and adiabatic cooling can later be reheated by subsequent hydrodynamic interaction, such as reflected shocks produced by the bipolar flow. 
This reheating can raise $T_{\rm e}$ and weaken or erase the condition $T_{\rm e}<T_Z$. 
Therefore, the visibility of recombining-plasma signatures depends not only on whether rapid adiabatic cooling occurred, but also on the later interaction history and the epoch at which the remnant is observed.

The evolution of $T_Z$ also provides a link to the $kT_{\rm init}$ parameter used in recombining-plasma spectral fits. 
Observationally, $kT_{\rm init}$ represents the effective high-temperature ionization state before rapid cooling. 
In our models, this corresponds most closely to the peak or pre-cooling value of $kT_Z$. 
For W49B, observations typically infer current electron temperatures of $kT_{\rm e}\sim 1$--$2$ keV and higher ionization or initial temperatures of order $kT_Z\sim 3$ keV or $kT_{\rm init}\sim 4$ keV for the recombining plasma \citep{ozawa2009suzaku, yamaguchi2018evidence}. 
The peak $kT_Z$ reached in the short-delay model, $\gtrsim 1$ keV, is lower than but within a factor of a few of the ionization temperatures inferred for W49B, suggesting that dense CSM interaction can raise Type Ia ejecta to an ionization state comparable to that inferred for W49B-like recombining plasma, although the present models are not intended as source-specific fits.

\subsection{X-ray properties}
\label{sec:x-ray}

\subsubsection{Volume-integrated spectra}

\begin{figure*}[ht!]
\includegraphics[width=0.95\textwidth]{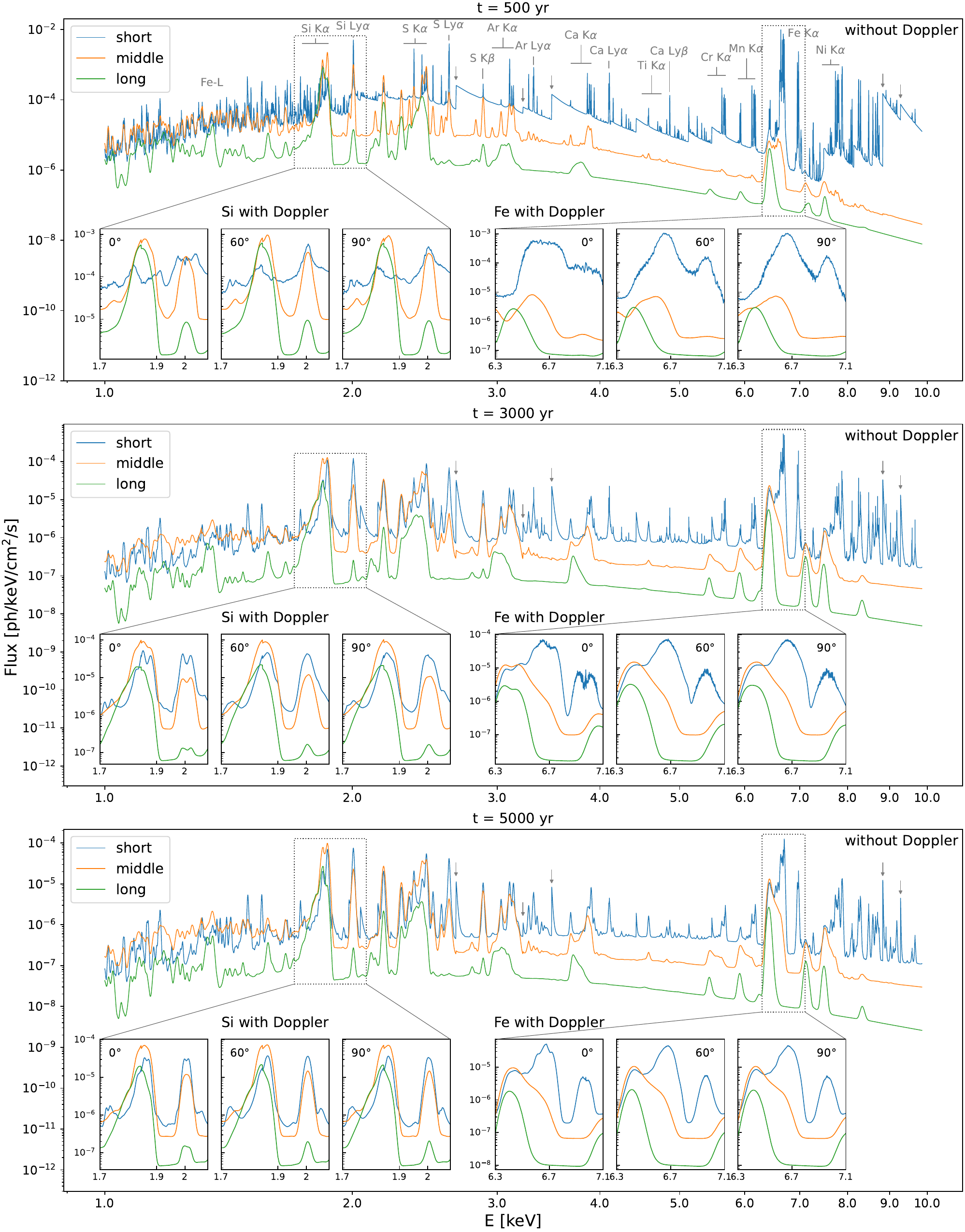}
\caption{Volume-integrated X-ray spectra of the shocked ejecta for three epochs.
Main panels show the spectra computed with thermal broadening only and without Doppler shifts. 
Inset panels show spectra including Doppler effects due to line-of-sight
velocities for three different line-of-sight directions.
The line-of-sight direction is specified by the angle between the line of sight
and the polar axis, such that $0^\circ$ and $90^\circ$ correspond to polar and
equatorial directions, respectively.
Colors indicate the PN models as labeled in the panels, and the rows from top to bottom show the time evolution. 
Labels indicate the principal K-shell emission-line complexes; the K$\alpha$ labels refer to the corresponding line complexes rather than single transitions. 
Arrows mark the radiative recombination continuum (RRC) edge energies, corresponding in increasing energy to Si~XIV, S~XV, S~XVI, Fe~XXV, and Fe~XXVI RRCs.
For the spectral calculation, we assume a distance of $11.3\,\mathrm{kpc}$ and a column density of $N_{\mathrm{H}}=5.0\times10^{22}\,\mathrm{cm}^{-2}$. The fluxes are not rescaled, so their differences primarily reflect luminosity differences among models and epochs.
\label{fig:spec}}
\end{figure*}

Figure~\ref{fig:spec} shows the volume-integrated X-ray spectra of the shocked ejecta at three representative epochs.
These spectra should be interpreted as spatially integrated spectra in which the overionized, adiabatically cooled, and reheated ejecta components discussed above are mixed along the line of sight and over the whole remnant.
Because these spectra in the main panels include thermal broadening but not Doppler shifts, they more directly trace differences in the underlying plasma conditions.
For the Doppler-shifted inset spectra, we define $\theta$ as the angle between the line of sight and the polar axis, with $\phi = 0^\circ$. 
We show $\theta = 0^\circ$, $60^\circ$, and $90^\circ$ as representative polar, intermediate, and equatorial viewing directions, respectively.
We note that, under the assumption of simple homologous expansion, the bipolar axis of W49B was inferred to be inclined by approximately $14^\circ$ from the plane of the sky \citep{xrism2025kinematic}.
This corresponds to $\theta \simeq 76^\circ$ in our definition, where $\theta$ is measured between the polar axis and the line-of-sight.

Across all epochs, the short-delay model is the brightest, whereas the long-delay model is the faintest. 
This trend reflects the stronger interaction in the short-delay case, where the ejecta encounter a denser and more compact CSM and therefore produce more X-ray-emitting plasma.

The short-delay model is characterized by relatively narrow lines and prominent radiative recombination continua (RRCs)\footnote{Here, ``radiative'' denotes the atomic recombination process rather than radiative cooling in the hydrodynamical evolution.}, especially at early times. 
An RRC appears as a continuum edge at the recombination threshold energy, followed by an approximately exponential decline toward higher energies, with a characteristic width of order $kT_e$. 
Thus, a sharp RRC feature indicates recombining plasma with a relatively low electron temperature.
These features indicate rapidly cooled recombining plasma produced by strong interaction with the dense CSM followed by rapid expansion. 
This picture is consistent with Figure~\ref{fig:Te_TCIE_Fe}, where the short-delay model reaches the $T_e<T_\mathrm{Z}$ regime, and with Figure~\ref{fig:Te_TCIE_dist_Fe}, where overionized plasma is concentrated in the inner ejecta and bipolar outflow. 
In this case, the electron temperature drops below that expected from the ionization state under the assumption of CIE, leading to recombination signatures. 
At 3000 and 5000 yr, the short-delay model also shows broader line components, which are likely associated with plasma reheated by reflected shocks in the bipolar structure, again consistent with the outer reheated component seen in Figure~\ref{fig:Te_TCIE_dist_Fe}.
At these late epochs, some asymmetric line-like features that cannot be readily identified with specific bound--bound transitions may in fact be narrow RRCs produced by low-$T_e$ recombining ejecta. 
An RRC has a sharp edge on the low-energy side and an exponential tail toward higher energies; as $T_e$ decreases, the tail becomes narrower and the feature can resemble a discrete emission line.
The low-$T_e$ component seen at late times in Figure~\ref{fig:Tion_Te_Fe} supports this interpretation.

The middle-delay model shows weaker recombining-plasma features than the short-delay model. 
Although weak RRC features can still be seen at 500 yr, they are much less prominent, indicating more moderate cooling and recombination. 
This is consistent with the weaker temperature--ionization inversion in Figure~\ref{fig:Te_TCIE_Fe}.

The long-delay model is dominated by broader lines and more weakly modified NEI plasma. 
Its line complexes generally correspond to lower ionization states than in the short-delay model, making the spectrum more similar to that of a standard Type Ia SNR evolving in a uniform-ISM medium. 
The differences in the line complexes should not be interpreted simply as differences in elemental abundance or total luminosity. 
They also reflect changes in the ionization populations. 
In lower-ionization plasma, the relative strengths of K-shell transitions change; in particular, the relative strengths of the H-like Ly$\alpha$ and He-like K$\alpha$ lines, particularly for IMEs such as Si and S differ from those in more highly ionized plasma. 
Thus, the apparent prominence of some K$\alpha$ features mainly traces the lower ionization state rather than a larger elemental mass. 
For Fe, the ionization state also affects the detailed structure and centroid of the Fe K complex, as discussed below.

From the short- to the middle- and long-delay models, the narrow line structures and RRC features become progressively less prominent.
This trend reflects the decreasing emissivity contributions from the low-$T_{\rm ion}$ component to the narrow line structures and the low-$T_e$, overionized component to the RRC features.
The EM-weighted temperatures in Figure~\ref{fig:Te_TCIE_Fe} should not be interpreted as representative temperatures of the ejecta. 
Instead, the volume-integrated spectrum is determined by the local distributions of $T_e$ and $T_{\rm ion}$ shown in Figure~\ref{fig:Tion_Te_Fe}, the spatial distribution of overionized plasma shown in Figure~\ref{fig:Te_TCIE_dist_Fe}, and the relative emissivity contributions of these components. 
In the main panels of Figure~\ref{fig:spec}, where Doppler shifts are not included, the widths of individual emission lines primarily reflect the temperatures of ions.
The prominence of narrow features in the volume-integrated spectrum therefore depends on the emissivity contribution of the low-$T_{\rm ion}$ component.
For a more direct comparison between the thermal components in Figure~\ref{fig:Tion_Te_Fe} and their contributions to the X-ray spectrum, we refer the reader to Appendix~\ref{sec:append_spec_comp}, where the short-delay ejecta are decomposed component by component and their corresponding spectra are shown.

From an observational point of view, the most important feature of the short-delay model is that RRC features remain visible even in the volume-integrated spectrum. 
This is qualitatively similar to W49B, where strong Fe RRC emission has been detected in spatially integrated X-ray spectra and interpreted as evidence for highly overionized plasma \citep{ozawa2009suzaku}. 
By contrast, the middle- and long-delay models show much weaker or no clear RRC features in the integrated spectra. 
These cases may be relevant to asymmetric Type Ia or Type Ia-candidate remnants in which strong CSM interaction shapes the morphology, but clear recombining-plasma signatures are not observed in the global spectrum.

The short-delay spectra also show that the integrated emission can contain both cool recombining plasma and hotter reheated components. 
This kind of multi-temperature structure is qualitatively reminiscent of recent high-resolution X-ray spectroscopy of N103B, where the spectrum requires multiple ejecta components including a high-temperature, highly ionized Fe-dominated plasma component \citep{holland2026xrism}. 
Although we do not fit our synthetic spectra with observational plasma models here, this comparison suggests that a dense, non-spherical CSM interaction can naturally produce integrated spectra in which low-temperature recombining features and hotter shocked-ejecta components coexist.

Thus, the PN delay time controls whether the volume-integrated spectrum appears as a bright recombining-plasma remnant, a more ordinary underionized Type Ia SNR, or an intermediate case in which asymmetric interaction is present but RRC features are weak.

\subsubsection{X-ray light curves}

\begin{figure*}[ht!]
\centering
\includegraphics[width=\textwidth]{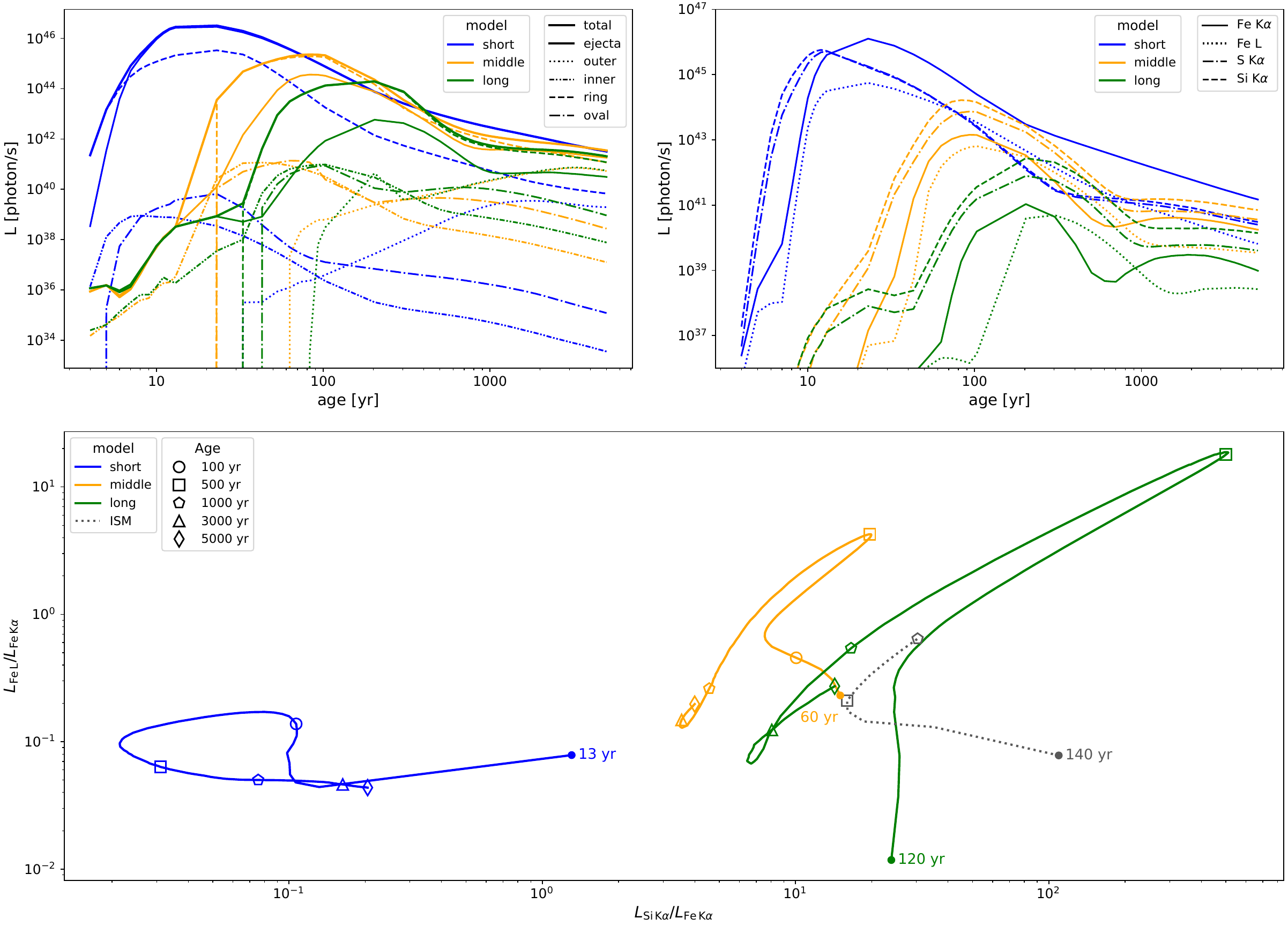}
\caption{Time evolution of the X-ray photon luminosity. The upper-left panel shows the total photon luminosity from different emitting components, while the upper-right panel shows the photon luminosities of selected lines (Fe K$\alpha$, Fe L, S K$\alpha$, and Si K$\alpha$) for the three PN models.
The bottom panel shows the evolution of the line-luminosity ratios Si K$\alpha$/Fe K$\alpha$ and Fe L/Fe K$\alpha$ , starting when the shocked fraction of the Fe ejecta reaches 10\%.
For comparison, the bottom panel shows the evolution of the uniform-ISM model up to 1000 yr.}
\label{fig:lc}
\end{figure*}

Figure~\ref{fig:lc} shows the time evolution of the X-ray photon luminosities. 
The left panel separates the emission by initial component, while the right panel shows selected line luminosities. 
These light curves provide a time-domain view of how the dominant emitting component changes during the ejecta--CSM interaction.

In the component-separated light curves, the initial brightening is delayed from the short- to the middle- and long-delay models. 
This reflects the increasing characteristic radius of the dense PN structure: the ejecta encounter the compact ring-like CSM earliest in the short-delay model and later in the longer-delay models. 
The short-delay model is dominated by shocked ejecta, and its luminosity evolves relatively smoothly after the early interaction. 
By contrast, the middle- and long-delay models show clearer rebrightening features. 
These rebrightenings likely reflect compression and reheating of the shocked plasma by reflected shocks, or renewed interaction with moderately dense ring-like CSM structures.
The dominant emitting component changes systematically with the CE-to-SN delay time. 
The X-ray emission is dominated mainly by the shocked ejecta in the short-delay model, while the ring contribution becomes increasingly important in the middle-delay model and dominates the emission in the long-delay model.

The contribution from the outer ambient medium remains small at early times, but increases once the forward shock reaches the outer region and continues to sweep up material. 
By $t\sim5000$ yr, the total luminosities of the three models become comparable. 
Thus, a similar late-time luminosity does not necessarily imply a similar interaction history.

The selected line light curves show that the emission history depends not only on the total amount of shocked material but also on the element and transition considered. 
The K$\alpha$ luminosities generally rise in the order Si, S, and Fe, reflecting the radial stratification of the ejecta: the outer IME-rich layers are shock-heated before the inner Fe-rich material. 
However, the relative behavior of the Fe and Si/S lines differs between the short-delay model and the longer-delay models, indicating that the line evolution is not determined by abundance stratification alone. 
It also depends on the different temperature and ionization histories produced by the CSM interaction.

The Fe L emission traces lower-energy Fe line complexes, whereas Fe K$\alpha$ is more sensitive to hotter and more strongly shocked Fe-rich plasma. 
The delayed evolution of the Fe L and Fe K$\alpha$ luminosities therefore reflects the later shock processing of the Fe-rich ejecta, together with the subsequent ionization and reheating history. 
In the middle- and long-delay models, the secondary brightenings seen in the Fe lines are likely associated with compression and reheating by reflected shocks or renewed interaction with moderately dense CSM structures.

Overall, the X-ray light curves show that the PN delay time controls not only when the remnant becomes bright, but also which physical component and which ejecta layer dominate the emission. 
A compact PN produces an early, ejecta-dominated bright phase, whereas more extended PN models produce delayed brightening, stronger rebrightening features, and a larger contribution from swept-up CSM and ambient material.

Finally, the line ratios shown in the bottom panels of Figure~\ref{fig:lc} provide a useful diagnostic of the ejecta--CSM interaction.
The Si K$\alpha$/Fe K$\alpha$ ratio reflects how deeply the ejecta have been processed by the reverse shock, because Si- and Fe-rich ejecta occupy different layers. 
As the reverse shock penetrates into the Fe-rich ejecta, the increasing Fe K$\alpha$ emission generally drives the ratio to lower values. 
At later times, however, the decreasing electron temperature suppresses Fe K$\alpha$ emission relative to Si K$\alpha$, causing the ratio to increase again. 
The Fe L/Fe K$\alpha$ ratio provides complementary information on the thermal and ionization state of the shocked Fe. 
In the short-delay model, the strong interaction with the compact CSM rapidly heats and ionizes the ejecta, producing relatively strong Fe K$\alpha$ emission and a smaller variation in the Fe L/Fe K$\alpha$ ratio. 
In the middle- and long-delay models, the weaker interaction results in a more pronounced evolution of the relative Fe L and Fe K$\alpha$ emission.
Therefore, these line ratios may provide a useful observational diagnostic of the CSM environment of Ia SNRs.
A more comprehensive investigation using other line ratios and direct comparisons with observed SNRs is beyond the scope of this work and will be explored in future studies.

\subsubsection{Region-specific spectra} \label{sec:region}

\begin{figure*}[ht!]
\includegraphics[width=\textwidth]{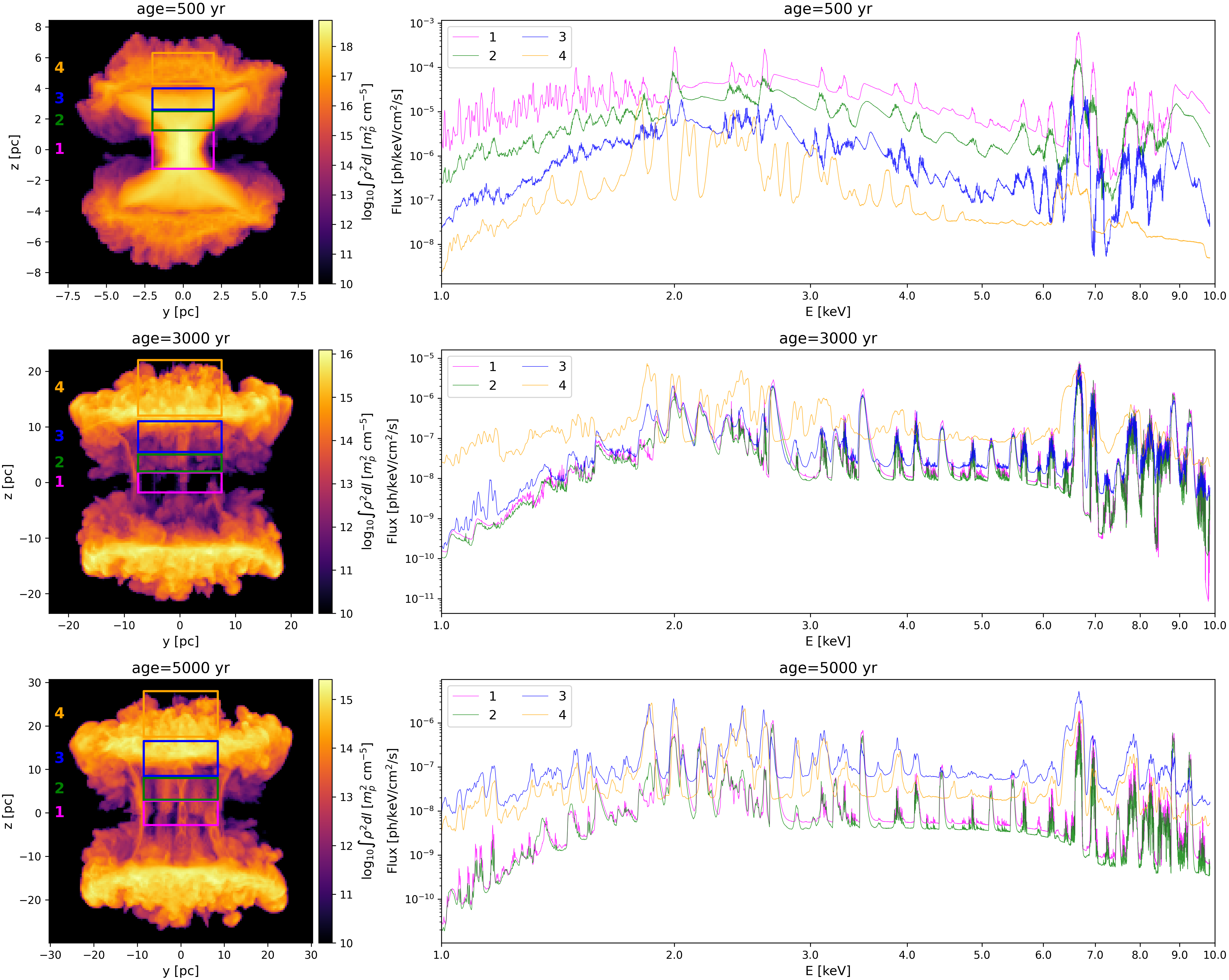}
\caption{Region-specific spectra and corresponding emission morphology for the short-delay model.
The line-of-sight direction corresponds to $\theta=90^\circ$ (observer on the equatorial plane).
The left panels show projected density-squared maps of ejecta, used as a proxy for the X-ray-emitting structure, constructed using shocked material selected by a pressure threshold. 
Boxes indicate the extraction regions for the spectra. 
The right panels show spectra of ejecta for each region, with colors corresponding to the same colored boxes in the left panels. 
These spectra are not rescaled and include both thermal broadening and Doppler shifts.
\label{fig:spec_4reg}}
\end{figure*}

Figure~\ref{fig:spec_4reg} shows region-specific spectra of the short-delay model.
The spectra are calculated for a viewing angle of $\theta=90^\circ$ for which the polar axis lies in the plane of the sky and the line-of-sight component of the large-scale bipolar flow is smaller than in more pole-on views.
The purpose of this analysis is to separate the plasma components that are mixed in the volume-integrated spectra shown in Figure~\ref{fig:spec}.
Unlike the main panels of Figure~\ref{fig:spec}, the region-specific spectra in Figure~\ref{fig:spec_4reg} include both thermal broadening and Doppler shifts.

The extraction regions are defined on the projected density-squared map, which is used as a proxy for the X-ray-emitting structure, with guidance from the $T_\mathrm{Z, Fe}$--$T_e$ distribution shown in Figure~\ref{fig:Te_TCIE_dist_Fe}. 
At each epoch, regions 1, 2, 3 and 4 are labeled from the center outward.
Region 1 samples the central ejecta, region 2 an off-center inner ejecta region, region 3 the interior of the upper bipolar outflow where rapidly expanded and cooled ejecta are prominent, and region 4 the outer polar shell where reheated ejecta make a larger contribution.

At early times, the central and interior regions make substantial contributions among the selected regions, while the shell contribution becomes progressively more important as the remnant evolves. 
This change in the dominant emitting region affects whether recombining-plasma signatures are visible in the volume-integrated spectra.
At the youngest epoch, the projected emission is relatively bright in the central and interior regions rather than being dominated by a limb-brightened shell. 
This center-brightened thermal morphology is reminiscent of mixed-morphology SNRs, although the present maps include only the shocked ejecta contribution and should not be interpreted as full broadband X-ray images.

At the youngest epoch, RRCs are visible in all selected regions except region 4. 
The sharp RRC features in regions 1, 2 and 3 indicate recombining plasma with a relatively low electron temperature, consistent with the rapid adiabatic cooling inferred from the 3D thermodynamic distribution.

By contrast, region 4 shows weaker recombining-plasma signatures and broader lines, indicating hotter plasma that has been reheated by reflected shocks. 
Thus, the same remnant contains both rapidly cooled recombining plasma and reheated shell plasma.

The Doppler behavior is also region dependent. 
Figure~\ref{fig:em_doppler_multiangle} shows projected density-squared maps of the shocked ejecta and the corresponding Doppler maps for the short-delay
model at $t=500$ and $5000$ yr for several viewing angles.
These maps connect the region-dependent line profiles in Figure~\ref{fig:spec_4reg} with the 3D velocity structure of the shocked ejecta. 
In region 1, Doppler substructure is relatively weak, whereas in region 3 the hourglass-like outflow has large line-of-sight velocity components, producing clear line splitting as seen in Figure~\ref{fig:spec_4reg}. 
Regions with clear RRC features tend to show narrower line widths, consistent with reduced thermal broadening in cooler plasma, while the reheated shell shows broader lines.

These results show that a young Type Ia SNR evolving inside a PN-like CSM can contain multiple plasma components with distinct X-ray spectral signatures. 
The visibility of recombining-plasma features therefore depends not only on the global plasma state, but also on the extraction region and the velocity structure along the line-of-sight.

\subsubsection{Doppler maps}

\begin{figure*}[ht!]
\centering
\includegraphics[width=0.91\textwidth]{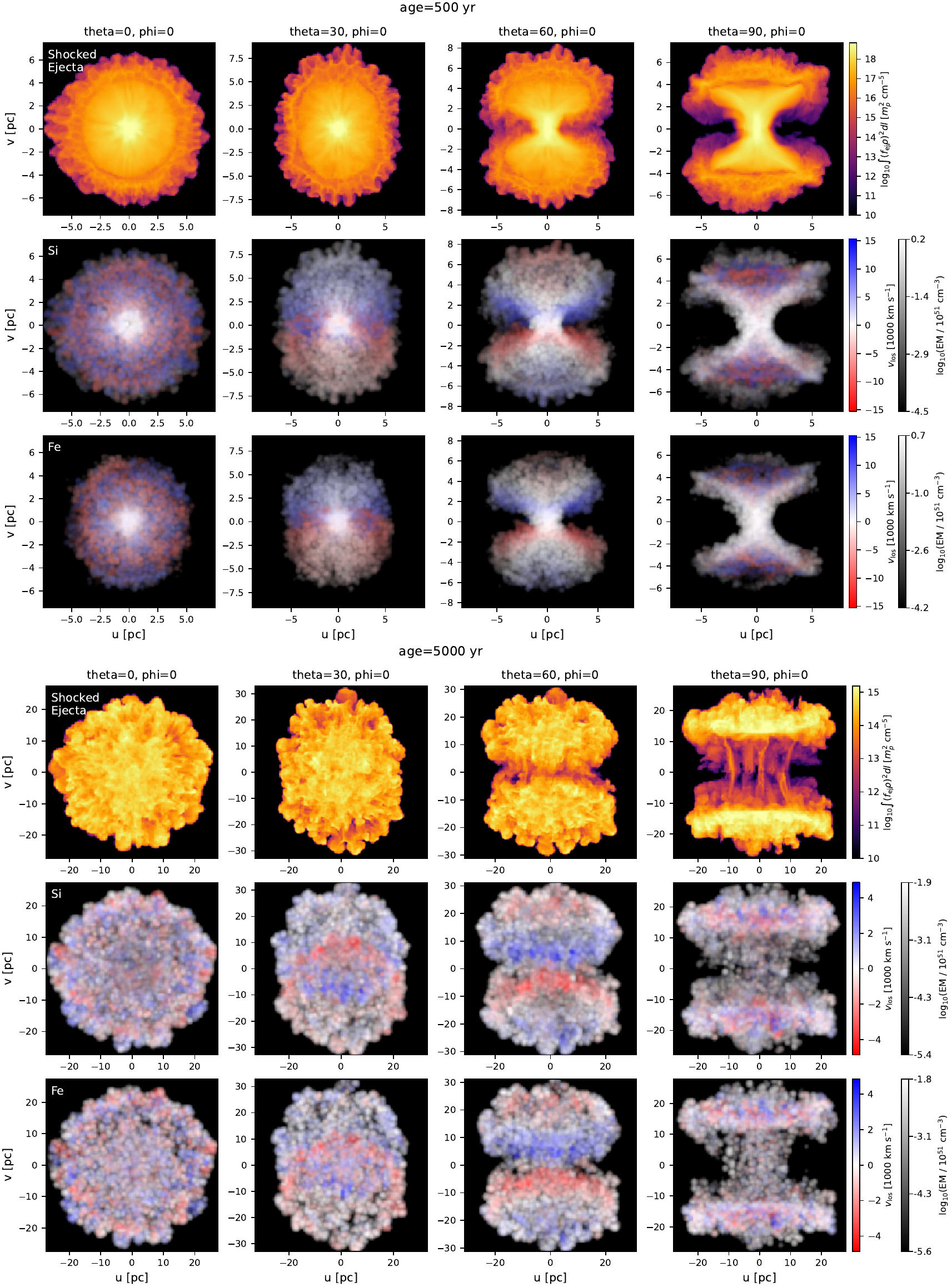}
\caption{Projected density-squared maps of shocked ejecta and corresponding Doppler maps for the short-delay model at two epochs. The upper and lower panels show the results at $t=500$ yr and $t=5000$ yr, respectively. The density-squared maps are constructed from ejecta identified as shocked material using a pressure threshold and a restricted spatial range, and therefore differ slightly from the emission maps shown in Figure~\ref{fig:spec_4reg}. The Doppler maps are shown for Si and Fe. In each projected pixel, the hue encodes the line-of-sight velocity distribution of the contributing particles: red indicates receding (redshifted) material, blue indicates approaching (blueshifted) material, white indicates material with small line-of-sight velocity, and purple indicates the coexistence of red- and blueshifted components. The transparency represents the emission measure (EM).}
\label{fig:em_doppler_multiangle}
\end{figure*}

Figure~\ref{fig:em_doppler_multiangle} illustrates how the 3D velocity field may be encoded in spatially resolved high-resolution X-ray spectra. 
The viewing angles of $\theta=0^\circ$, $60^\circ$, and $90^\circ$ correspond to those used for the Doppler-shifted inset spectra in Figure~\ref{fig:spec}, while $\theta=30^\circ$ is shown only in the Doppler maps.
In particular, coherent red- and blue-shifted regions may appear observationally as spatial variations of line centroids, line broadening, or line splitting. 
Thus, the Doppler maps provide an interpretive link between spectroscopic signatures and the underlying ejecta dynamics, although they should be regarded as qualitative emission-weighted velocity maps rather than full mock observations.

In the Doppler maps, red and blue indicate receding and approaching components, respectively. 
Purple regions indicate the coexistence of redshifted and blueshifted components along the same line of sight, rather than a single 
intermediate-velocity component. 
The color pattern therefore reflects both the large-scale velocity field and local emission-measure enhancements along each line of sight.

The most prominent large-scale feature is the bipolar velocity structure. 
At $t=500$ yr, the shocked ejecta are redirected along the bipolar axis by the interaction with the dense equatorial CSM, producing coherent approaching and receding components depending on viewing angle. 
This global bipolar flow explains the region-dependent line profiles in Figure~\ref{fig:spec_4reg}. 
The central extraction region contains approaching and receding components that are relatively weak or nearly symmetric, so clear line splitting is not prominent. 
By contrast, the upper cooling region overlaps with the hourglass-like bipolar flow, where multiple line-of-sight velocity components coexist. 
This produces the split line profiles seen in the corresponding regional spectrum.

In addition to this global bipolar pattern, the Doppler maps show small-scale red and blue patches. 
These local color variations should not be interpreted as coherent one-sided bulk motion of the remnant. 
Rather, they indicate that clumpy ejecta structures can locally dominate the emission measure along a given line of sight, making either the approaching or receding component appear stronger in projection. 
Such clumpiness may be produced by hydrodynamic instabilities during the SNR evolution and may also reflect intrinsic asymmetries in the explosion ejecta. 
Similar clumpy structures are also present in our previous simulations of SNRs evolving in a uniform-ISM \citep{fujimaru2026probing}, although they were not discussed there in detail.
On large scales, however, the bipolar ejecta contain both approaching and receding components.

At $t=5000$ yr, ring-like ejecta structures associated with the bipolar flow become particularly apparent. 
These should be distinguished from the initial equatorial CSM ring; they are ejecta structures shaped by the bipolar expansion and subsequent interaction. 
Their Doppler appearance depends strongly on viewing angle. 
For example, at $\theta=30^\circ$, the projected image shows polar ring-like structures in which approaching and receding components are both visible. 
At $\theta=60^\circ$, the rings are viewed more nearly from the side, so the two polar structures show clearer blue- and red-shifted components. 
This viewing-angle dependence indicates that ring-like projected morphologies and their Doppler signatures can arise naturally from a 3D bipolar ejecta structure.

At late times, the shocked ejecta still show structured emission in the projected maps, but the absolute density-squared values are much lower than at $t=500$ yr. 
Moreover, the total X-ray emission at this stage is increasingly affected by shocked CSM and swept-up ambient material. 
Thus, the apparent ejecta structure in this figure should not be interpreted as the total X-ray morphology of the remnant at late times.

\subsubsection{\texorpdfstring{Iron K$\alpha$ line diagnostics}{Iron K alpha line diagnostics}} \label{sec:yamaguchi}

\begin{figure*}[ht!]
\includegraphics[width=\textwidth]{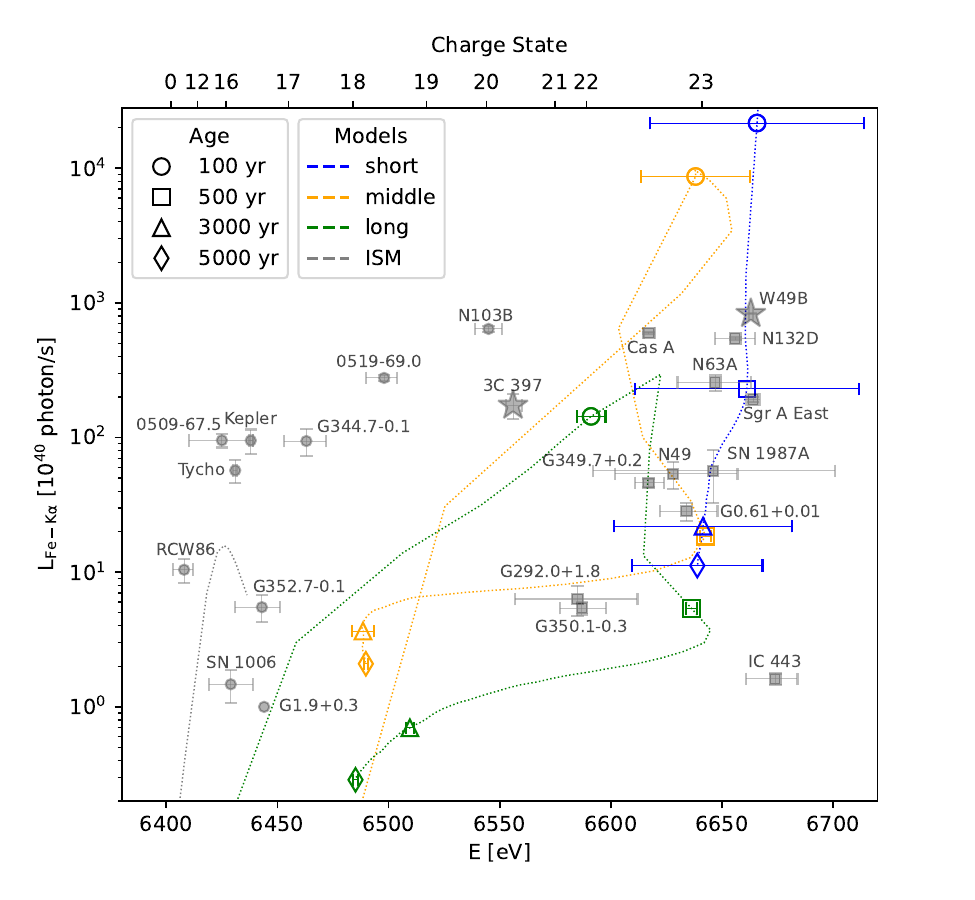}
\caption{Centroid energy and line luminosity of Fe K$\alpha$ for the SNR models at different ages, compared with observed Type Ia and CC SNRs \citep{yamaguchi2014discriminating}. 
The colored symbols show theoretical values computed from the Fe K$\alpha$ only; nearby line blends, such as Mn K$\beta$, are excluded. 
The gray circles and squares denote observed Type Ia and CC SNRs, respectively. 
The stars mark W49B and 3C 397, which are shown separately because their progenitor classifications remain controversial \citep[for abundance analyses, see, e.g.,][]{zhou2018asymmetric,treyturik2026revisiting}.
The plotted model values do not include Doppler shifts from bulk motion, while the error bars show the maximum possible centroid shifts caused by Doppler motion over all viewing angles.
\label{fig:yamaguchi}}
\end{figure*}

Figure~\ref{fig:yamaguchi} shows the Fe K$\alpha$ centroid energy and line luminosity in the diagnostic plane introduced by \citet{yamaguchi2014discriminating}. 
In this plane, the centroid energy mainly traces the ionization state of Fe, whereas the line luminosity reflects the amount of Fe K$\alpha$-emitting plasma and therefore the strength of the ejecta--ambient-medium interaction.

For the model points, we plot theoretical Fe K$\alpha$ quantities computed directly from the simulated plasma state, including the density, electron temperature, and ionization state, rather than values obtained by fitting the Fe K$\alpha$ complex in the synthetic spectra. 
Only the Fe K$\alpha$ emissivity is included when calculating the centroid energy and line luminosity; nearby blends, such as Mn K$\beta$, are excluded. 
The error bars show the possible centroid shifts caused by bulk Doppler motion over different viewing angles, evaluated using the same line-of-sight direction sampling method as in \citet{fujimaru2026probing}.

The exclusion of nearby blends is more important for the present PN models than for the uniform ambient-medium models of \citet{fujimaru2026probing}. 
In the latter case, the ionization states were generally lower, so the contribution of nearby Mn K$\beta$ emission to the Fe K$\alpha$ was weak and had little effect on the inferred Fe K$\alpha$ centroid. 
In the present models, especially in the short-delay case, the higher ionization state can make such nearby transitions more prominent; including them would bias the Fe K$\alpha$ centroid toward lower energies.

Whereas the uniform-ISM models in \citet{fujimaru2026probing} occupy only the lower-ionization region associated with observed Type Ia remnants, the short-delay models extend into the more highly ionized region where many CC SNRs are found. 
This behavior is a natural consequence of the compact and dense PN in the short-delay case, which drives rapid early ionization of the ejecta through stronger interaction. 
The higher plasma density produces a larger ionization state, shifting the Fe K$\alpha$ centroid to higher energies. 
It also increases the line luminosity because the emissivity scales with the emission measure, $EM = \int n_e n_{\mathrm{ion}}\, dV$ as in Equation~(\ref{eq:EM}).

By contrast, the middle- and long-delay models remain in the lower-ionization region, similarly to the uniform-ISM models of \citet{fujimaru2026probing}. 
The uniform-ISM model shown here corresponds to the \texttt{N100ddt} model with $n_{\rm ISM}=0.3 \,\mathrm{cm}^{-3}$ in \citet{fujimaru2026probing}. 
Although some of the other explosion models explored in \citet{fujimaru2026probing} reached higher Fe~K$\alpha$ centroid energies, the centroids without Doppler shifts remained below $\sim6.5$ keV.
Their PN-like CSM are more extended and lower density than in the short-delay case, so the ejecta are ionized less efficiently and the Fe K$\alpha$ emission remains fainter. 
In addition, because the surrounding ISM density adopted in the present models is lower than in \citet{fujimaru2026probing}, our models generally show both lower ionization and lower luminosity than those uniform-medium calculations. 
Part of the luminosity difference also reflects the absorption treatment adopted in the present analysis: the fiducial absorbing column density used here, $N_{\rm H}=5.0\times10^{22}\ {\rm cm^{-2}}$, makes the plotted Fe K$\alpha$ luminosities slightly fainter than they would be for a less absorbing setup.

This behavior can be understood by comparing our results with the 1D PN calculations of \citet{Court_2024}.
In their equatorial models, the dense PN component effectively acts as a spherically covering dense wall: the ejecta have no lateral escape route, and a large fraction of the shocked ejecta experiences rapid heating and ionization.
This naturally produces high Fe K$\alpha$ centroid energies and bright Fe K$\alpha$ emission.
Our short-delay model shows a qualitatively similar trend, suggesting that a compact and dense equatorial ring can dominate the thermal and ionization history of the X-ray-bright ejecta even in three dimensions.

The middle- and long-delay models, however, do not simply follow the corresponding 1D equatorial behavior.
In our 3D geometry, the dense material occupies only a limited solid angle, and the shocked ejecta can be redirected toward lower-density polar regions rather than being forced to interact with a dense wall everywhere.
The fraction of ejecta that undergoes rapid ionization in dense CSM, and the subsequent EM contribution of that plasma, are therefore controlled not only by the density of the CSM but also by the covering and geometry of the ring and by the post-shock expansion of the redirected flow.
This is a genuinely multidimensional effect: a compact dense ring can produce highly ionized Fe emission, whereas a more extended or less effectively covering ring can leave the remnant closer to ordinary Type Ia SNRs in the Fe K$\alpha$ plane.

Overall, these results suggest that the Fe K$\alpha$ centroid energy and line luminosity do not map uniquely to the explosion mechanism itself; rather, they primarily trace the strength of the ejecta--ambient-medium interaction during the transition from SN to SNR. 
This view is consistent with the 1D Type Ia SNR study of \citet{martinez2018chandrasekhar}, which showed that Fe K$\alpha$ and other bulk X-ray properties are driven mainly by the ambient density and remnant age, rather than by the detailed differences among explosion models.
It is also consistent with the classification scheme of \citet{court2026classification}, in which X-ray-bright Type Ia SNRs are classified in terms of their circumstellar interaction using bulk properties including Fe K$\alpha$ centroid energies and luminosities.
Similar behavior is also seen in the 1D CC SNR simulations of \citet{kawashima2026single}.
In the PN models, this interaction strength is controlled by the characteristic CSM scale set by the delay time, implying that a single Type Ia origin can reproduce a wide range of plasma states observed in Type Ia SNRs simply through differences in interaction.
Thus, although this diagnostic remains useful for empirical Ia/CC classification, its physical meaning should be understood primarily as a tracer of interaction strength rather than as a direct signature of the explosion mechanism itself.
In this sense, the apparent separation between many Type Ia and CC remnants mainly reflects the tendency for Type Ia SNRs to undergo weaker interaction, whereas CC SNRs more often undergo stronger interaction, rather than indicating that the two classes are intrinsically cleanly separated in this plane.

\section{Summary} \label{sec:summary}

In this study, we have presented a test case for modeling the evolution of Type Ia supernova remnants in a fully 3D environment, from hydrodynamical evolution with NEI to synthetic X-ray spectra. 
As a representative example, we considered a Type Ia supernova exploding inside a PN-like CSM, motivated by a post-CE progenitor scenario.

The CSM was constructed as a set of simplified toy models based on the 2D PN simulations of \citet{garcia2018common}.
We considered three representative delay times after the CE phase, corresponding to different characteristic sizes and densities of the circumstellar structure. 
For the supernova ejecta, we adopted the near-Chandrasekhar-mass delayed-detonation model N100ddt of \citet{seitenzahl2013three}. 
The subsequent remnant evolution was calculated by extending the hydrodynamics and NEI framework of \citet{fujimaru2026probing} to the 3D circumstellar environment.

Within this framework, we examined how the characteristic scale of the PN affects the morphology, ionization state, and observable X-ray properties of the resulting Type Ia remnant. 
Our main results are summarized as follows.

\begin{itemize}
\setlength{\itemsep}{-5pt}

 \item The CE-to-SN delay time, defined as the interval between the ejection of the PN and the SN explosion, strongly controls the remnant morphology. 
 Shorter delay times produce smaller and denser PN-like CSM, leading to earlier and stronger ejecta--CSM interaction.
 This interaction suppresses equatorial expansion and drives bipolar expansion, producing asymmetric and ring-like ejecta structures even from the same Type Ia explosion model.

 \item Interaction with the dense CSM produces a distinctive plasma evolution.
 The shocked ejecta experience rapid heating and ionization during the early collision with the dense ring, followed by adiabatic cooling as the flow expands toward lower-density polar regions. 
 As a result, the models show thermal and ionization histories that differ from those of ordinary Type Ia SNRs evolving in a uniform ambient medium.

 \item The shock-processing history depends strongly on the delay time. 
 In the short-delay model, the reverse shock reaches the inner, Fe-rich ejecta at an early stage, while the shock processing proceeds more slowly in the middle- and long-delay models. 
 In particular, the long-delay model evolves more similarly to the fixed-density uniform-ISM case, because the PN-like CSM is more extended and has a lower density when the SN explodes.

 \item Signatures of overionized plasma appear locally in all PN-like CSM models. 
 They are produced by the combination of rapid ionization during the dense-ring interaction and subsequent adiabatic cooling in the polar expansion. 
 In the short-delay model, the overionization is strong enough to appear even in EM-averaged quantities, and recombining-plasma signatures such as radiative recombination continua are expected to be visible in the integrated X-ray spectrum. 
 The averaged ionization temperature in this model is also broadly consistent with that inferred for W49B.

 \item The synthetic X-ray observables show clear delay-time dependence. 
 The short-delay model is brighter in X-rays and reaches higher ionization states than the middle- and long-delay models.
 Its high-resolution spectra show both hot and cool plasma components, and localized spectra exhibit Doppler shifts associated with the bipolar expansion. 
 At early stages, the short-delay model also shows a centrally bright, mixed-morphology-like X-ray appearance.

 \item The Fe K$\alpha$ properties mainly trace the strength and history of the ejecta--CSM/ISM interaction, rather than the detailed properties of the explosion model, consistent with the 1D Type Ia SNR calculations of \citet{martinez2018chandrasekhar}.
 The short-delay model remains in the CC-like region of the Fe K$\alpha$ centroid--luminosity plane for much of its evolution, while the middle- and long-delay models are also shifted toward higher ionization and higher luminosity compared with uniform-ISM models.
 Although this plane provides a useful empirical separation between Type Ia and CC SNRs \citep{yamaguchi2014discriminating},
 our results indicate that Fe K$\alpha$ diagnostics largely reflect the different interaction histories \citep{court2026classification} and ambient environments typically associated with the two populations.
 Therefore, a Type Ia SNR evolving in dense, structured CSM can show CC-like Fe K$\alpha$ properties.
 If W49B is indeed a Type Ia SNR, it would represent an extraordinary example of such a strongly interacting Type Ia SNR.

 \item Overall, our results demonstrate that a single Type Ia explosion model can produce many different remnant appearances when it evolves inside PN-like CSM with different delay times. 
 Bipolar morphologies, ring-like ejecta structures, bright and highly ionized Fe~K emission, multi-temperature plasma, recombining-plasma signatures, and mixed-morphology-like X-ray emission can all arise from 3D interaction with PN-like CSM. 
 These results support an environmental origin for at least part of the observed diversity of Type Ia and Type Ia-candidate SNRs.

\end{itemize}

The present work extends the framework of \citet{fujimaru2026probing} from a uniform ambient medium to a fully 3D, non-spherical CSM.
By combining 3D hydrodynamical evolution, NEI calculations, and synthetic X-ray spectra, we have shown that a Type Ia explosion model can produce large-scale asymmetric morphologies, in particular bipolar and ring-like ejecta structures, when it evolves inside PN-like CSM.
The same interaction also drives adiabatic expansion cooling in 3D, producing signatures of overionized plasma that appear in both line emission and continuum features.

In the present study, the equatorial CSM mass is fixed at $1 \,M_\odot$, while the characteristic scale and density are varied through the CE-to-SN delay time. 
The resulting differences among the models indicate that the impact of the CSM depends on how compact the material is at the time of explosion. 
However, the combined dependence on CSM mass and spatial scale remains to be explored, including whether lower-mass CSM can still significantly affect the SNR evolution when located sufficiently close to the explosion. 

This approach is useful for interpreting high-resolution X-ray observations.
Recent XRISM/Resolve observations have revealed complex plasma and velocity structures in remnants such as W49B and N103B, including bar-like kinematics or multiple thermal components \citep[e.g.,][]{xrism2025kinematic, holland2026xrism}.
In Sgr~A East, for which a Type Iax-like thermonuclear origin has been proposed, XRISM/Resolve measurements of the resolved Fe~XXV He$\alpha$ complex, particularly the forbidden-to-resonance line ratio, provide strong evidence for overionized plasma \citep{xrism2025overionized}.
Our results suggest that such features can arise, at least in part, from 3D interaction between Type Ia ejecta and structured PN-like CSM.

Several limitations of our models should also be noted.
We adopt the relatively spherical \texttt{N100ddt} explosion model to isolate the effects of CSM interaction, whereas companion interactions may introduce additional asymmetries in the ejecta.
Furthermore, a core-degenerate progenitor may be more naturally associated with the PN-like environment considered here.
Different progenitor and explosion models could produce different ejecta structures and nucleosynthetic yields, resulting in different X-ray spectral signatures, as demonstrated for different explosion models by \citet{fujimaru2026probing}.
Realistic PNe may also contain complex internal density structures and 3D asymmetries, including differences between the two polar lobes, which are not represented in our simplified CSM models.
In addition, we assume solar abundances for the CSM, although the chemical composition of PN material may be altered by preceding stellar evolution.
These effects could introduce additional morphological and spectral diversity beyond that explored in the present study.

Therefore, future studies should explore a wider parameter space, including different CE-to-SN delay times, PN-like CSM geometries, ambient environments, and explosion models.
Detailed comparisons of line widths, Doppler shifts, ionization states, and radiative recombination continua will help constrain not only the explosion mechanism, but also the circumstellar environment in which Type Ia SNRs evolve.
Such 3D modeling will also be important for interpreting future spatially resolved, high-resolution X-ray spectroscopic observations of SNRs with NewAthena \citep{cruise2025newathena}, planned for the 2030s.

\begin{acknowledgments}
 We are grateful to Jacco Vink and Manami Sasaki for helpful discussions.
 This work was supported by JST SPRING, Grant Number JPMJSP2110. 
 H.L. acknowledges support from JSPS grant No. JP19K03913. 
 C.B. acknowledges support from Chandra Theory grant Nos. TM0-21004X and TM1-22004X, and XRISM Guest Scientist grant No. 80NSSC23K0634. 
 K.M. acknowledges support from JSPS KAKENHI grant (JP26K22330, JP24KK0070, and JP23H04894). 
 S.N. is supported by JSPS Grant-in-Aid Scientific Research (KAKENHI) (A), Grant Number JP25H00675, and (B), Grant Number JP23K25874, and JST ASPIRE Program “RIKEN-Berkeley mathematical quantum science initiative”. 
 D.J.P. acknowledges support from the Chandra X-ray Center, which is operated by the Smithsonian Institution under NASA contract NAS8-03060. 
 S.S.H. acknowledges support from the Natural Sciences and Engineering Research Council of Canada (NSERC) through the Canada Research Chairs and Discovery Grants Programs. 
 Supercomputing support in Canada is provided by the Digital Research Alliance of Canada. 
 The work of F.K.R. is supported by the Klaus Tschira Foundation, the Deutsche Forschungsgemeinschaft (DFG, German Research Foundation) – RO 3676/7-1, project number 537700965, and by the European Union (ERC, ExCEED, project number 101096243). 
 Views and opinions expressed are, however, those of the authors only and do not necessarily reflect those of the European Union or the European Research Council Executive Agency. 
 Neither the European Union nor the granting authority can be held responsible for them. 
 A.D. acknowledges financial support from the Centre national d’études spatiales (CNES), France (ROR: https://ror.org/04h1h0y33) within the framework of the XRISM/NewAthena missions.
\end{acknowledgments}

\appendix

\section{spectral components}
\label{sec:append_spec_comp}

\renewcommand{\thefigure}{A\arabic{figure}}
\setcounter{figure}{0}

\begin{figure*}[ht!]
\includegraphics[width=\textwidth]{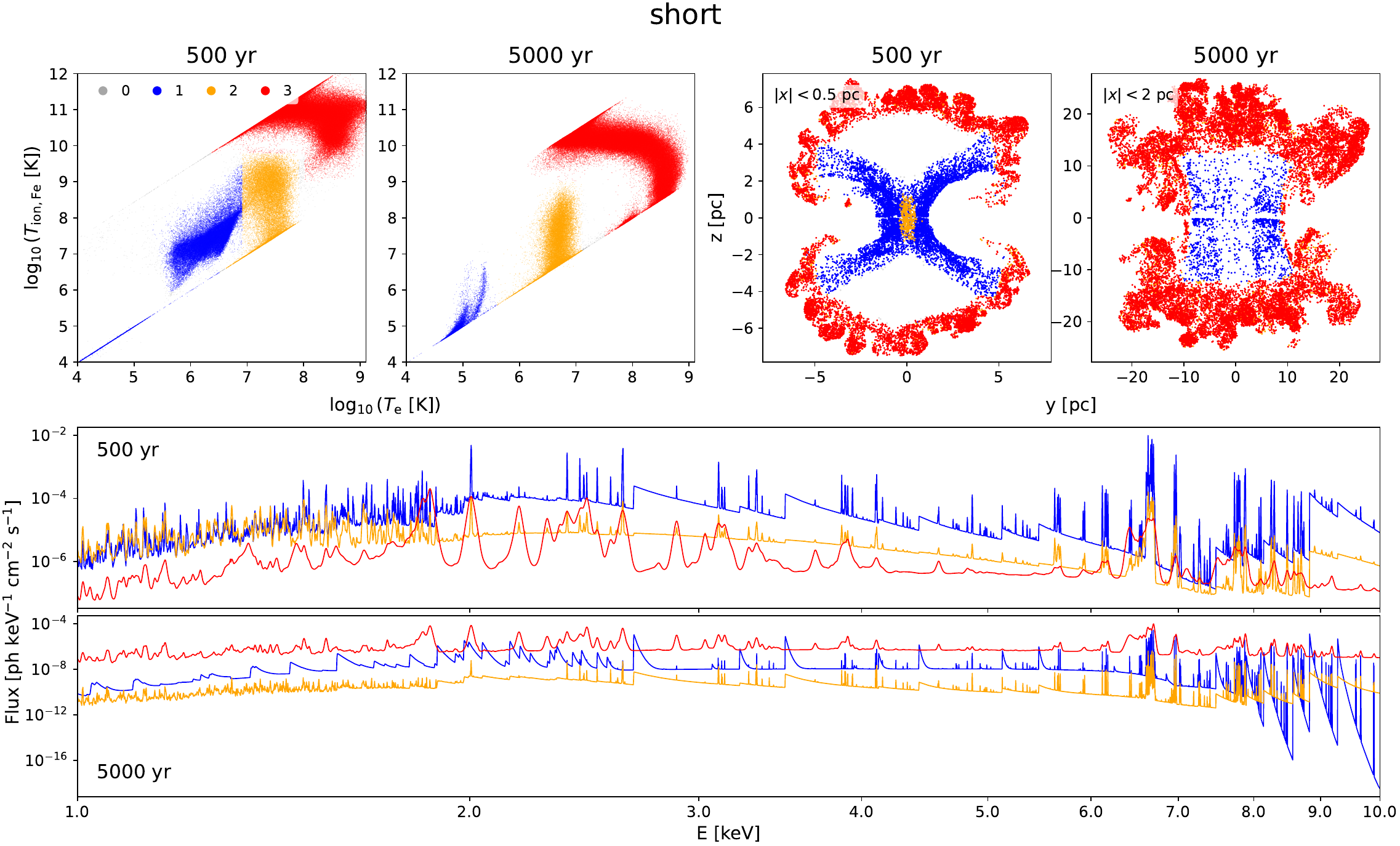}
\caption{Comparison of the distinct ejecta components in the $T_{\mathrm e}$–$T_{\mathrm{ion,Fe}}$ plane, their spatial distributions, and their synthetic X-ray spectra. 
The top-left panels show individual ejecta tracer particles in the same temperature plane as Figure~\ref{fig:Tion_Te_Fe}, colored according to their ejecta components.
Gray indicates particles not assigned to any of the three groups, which are not included in the spectra shown in the bottom panels.
The top-right panels show slice-like maps of the ejecta-component distributions near the (x=0) plane. These maps have a finite thickness and therefore do not represent exact planar slices.
The bottom panels show the corresponding synthetic X-ray spectra, including thermal broadening but excluding Doppler shifts.
\label{fig:spec_comp_temp}}
\end{figure*}

Figure~\ref{fig:spec_comp_temp} shows the distribution of distinct ejecta components in the $T_{\mathrm e}$–$T_{\mathrm{ion,Fe}}$ plane, their spatial distributions, and their corresponding X-ray spectra. 
We use this decomposition to identify the origin of the narrow spectral features seen in the integrated spectrum. 
The component-resolved spectra differ substantially in both line width and intensity. 
Because only thermal broadening is included, the line widths reflect the thermal velocities of the emitting ions, while the prominence of each feature in the integrated spectrum depends on the relative emission strength of the corresponding component. 
The low-temperature recombining component produces the narrowest features. 
Spatially, this component is associated with ejecta that interact with the ring near the center and subsequently flow away from the interaction region, where they cool as they expand. 
These results indicate that the narrow features in the integrated spectrum originate primarily from the low-temperature recombining ejecta.

\bibliography{references}{}

@ARTICLE{kwok2026ApJ..1005...52K,
       author = {{Kwok}, Lindsey A. and {Liu}, Chang and {Jha}, Saurabh W. and {Blondin}, St{\'e}phane and {Larison}, Conor and {Miller}, Adam A. and {Dai}, Mi and {Foley}, Ryan J. and {Filippenko}, Alexei V. and {Andrews}, Jennifer E. and {Andrews}, Moira and {Auchettl}, Katie and {Badenes}, Carles and {Bostroem}, K. Azalee and {Brink}, Thomas G. and {Koelln}, Cristine and {Davis}, Kyle W. and {Fl{\"o}rs}, Andreas and {Galbany}, Llu{\'\i}s and {Graur}, Or and {Howell}, D. Andrew and {Kumar}, Sahana and {K{\"o}nyves-T{\'o}th}, R{\'e}ka and {LeBaron}, Natalie and {Macrie}, Colin W. and {Maeda}, Keiichi and {Maguire}, Kate and {McCully}, Curtis and {Meza-Retamal}, Nicolas E. and {Padilla Gonzalez}, Estefania and {Pakmor}, R{\"u}diger and {Pearson}, Jeniveve and {Piro}, Anthony L. and {Polin}, Abigail and {Rehemtulla}, Nabeel and {Rojas-Bravo}, C{\'e}sar and {Sand}, David J. and {Sangkachan}, Chita and {Schwab}, Michaela and {Sears}, Huei and {Singh}, Mridweeka and {Subrayan}, Bhagya M. and {Taggart}, Kirsty and {Temim}, Tea and {Terwel}, Jacco H. and {Tinyanont}, Samaporn and {Vink{\'o}}, J{\'o}zsef and {Wang}, Xiaofeng and {Wheeler}, J. Craig and {Yang}, Yi and {Zheng}, WeiKang},
        title = "{JWST Spectroscopy of SNe Ia 2022aaiq and 2024gy: Evidence for Enhanced Central Stable Ni Abundance and a Deflagration-to-detonation Transition}",
      journal = {\apj},
         year = 2026,
        month = jul,
       volume = {1005},
       number = {1},
          eid = {52},
        pages = {52},
          doi = {10.3847/1538-4357/ae6867},
       adsurl = {https://ui.adsabs.harvard.edu/abs/2026ApJ..1005...52K}
}

@ARTICLE{katsuda2015ApJ...808...49K,
       author = {{Katsuda}, Satoru and {Mori}, Koji and {Maeda}, Keiichi and {Tanaka}, Masaomi and {Koyama}, Katsuji and {Tsunemi}, Hiroshi and {Nakajima}, Hiroshi and {Maeda}, Yoshitomo and {Ozaki}, Masanobu and {Petre}, Robert},
        title = "{Kepler{\textquoteright}s Supernova: An Overluminous Type Ia Event Interacting with a Massive Circumstellar Medium at a Very Late Phase}",
      journal = {\apj},
         year = 2015,
        month = jul,
       volume = {808},
       number = {1},
          eid = {49},
        pages = {49},
          doi = {10.1088/0004-637X/808/1/49},
archivePrefix = {arXiv},
       eprint = {1506.03135},
 primaryClass = {astro-ph.HE},
       adsurl = {https://ui.adsabs.harvard.edu/abs/2015ApJ...808...49K}
}

@article{treyturik2026revisiting,
          doi = {10.3847/1538-4357/ae2c50},
          url = {https://doi.org/10.3847/1538-4357/ae2c50},
         year = {2026},
        month = {jan},
    publisher = {The American Astronomical Society},
       volume = {997},
       number = {2},
        pages = {211},
       author = {Treyturik, Cole and Braun, Chelsea and Safi-Harb, Samar and Fryer, Christopher L. and Ferrand, Gilles},
        title = {Revisiting the Supernova Engines in the 3C 397 and W49B Supernova Remnants},
      journal = {The Astrophysical Journal}
}

@article{ferrand2019supernova,
        title = {From supernova to supernova remnant: the three-dimensional imprint of a thermonuclear explosion},
       author = {Ferrand, Gilles and Warren, Donald C and Ono, Masaomi and Nagataki, Shigehiro and R{\"o}pke, Friedrich K and Seitenzahl, Ivo R},
      journal = {The Astrophysical Journal},
       volume = {877},
       number = {2},
        pages = {136},
         year = {2019},
    publisher = {IOP Publishing},
          url = {https://iopscience.iop.org/article/10.3847/1538-4357/ab1a3d},
          doi = {10.3847/1538-4357/ab1a3d}
}

@article{pakmor2022fate,
        title = {On the fate of the secondary white dwarf in double-degenerate double-detonation Type Ia supernovae},
       author = {Pakmor, R{\"u}diger and Callan, FP and Collins, CE and de Mink, SE and Holas, A and Kerzendorf, WE and Kromer, M and Neunteufel, PG and O’Brien, John T and R{\"o}pke, FK and others},
      journal = {\mnras},
       volume = {517},
       number = {4},
        pages = {5260--5271},
         year = {2022},
    publisher = {Oxford University Press},
          url = {https://academic.oup.com/mnras/article/517/4/5260/6779709},
          doi = {10.1093/mnras/stac3107}
}

@article{yamaguchi2014discriminating,
        title = {Discriminating the progenitor type of supernova remnants with iron K-shell emission},
       author = {Yamaguchi, Hiroya and Badenes, Carles and Petre, Robert and Nakano, Toshio and Castro, Daniel and Enoto, Teruaki and Hiraga, Junko S and Hughes, John P and Maeda, Yoshitomo and Nobukawa, Masayoshi and others},
      journal = {The Astrophysical Journal Letters},
       volume = {785},
       number = {2},
        pages = {L27},
         year = {2014},
    publisher = {IOP Publishing},
          url = {https://iopscience.iop.org/article/10.1088/2041-8205/785/2/L27},
          doi = {10.1088/2041-8205/785/2/L27}
}

@article{seitenzahl2013three,
        title = {Three-dimensional delayed-detonation models with nucleosynthesis for Type Ia supernovae},
       author = {Seitenzahl, Ivo R and Ciaraldi-Schoolmann, Franco and R{\"o}pke, Friedrich K and Fink, Michael and Hillebrandt, Wolfgang and Kromer, Markus and Pakmor, R{\"u}diger and Ruiter, Ashley J and Sim, Stuart A and Taubenberger, Stefan},
      journal = {Monthly Notices of the Royal Astronomical Society},
       volume = {429},
       number = {2},
        pages = {1156--1172},
         year = {2013},
    publisher = {The Royal Astronomical Society},
          url = {https://academic.oup.com/mnras/article/429/2/1156/1035579},
          doi = {10.1093/mnras/sts402},
       eprint = {https://academic.oup.com/mnras/article-pdf/429/2/1156/18453705/sts402.pdf}
}

@article{blondin2001rayleigh,
        title = {Rayleigh-Taylor instabilities in young supernova remnants undergoing efficient particle acceleration},
       author = {Blondin, John M and Ellison, Donald C},
      journal = {The Astrophysical Journal},
       volume = {560},
       number = {1},
        pages = {244},
         year = {2001},
    publisher = {IOP Publishing},
          url = {https://iopscience.iop.org/article/10.1086/322499},
          doi = {10.1086/322499}
}

@article{zuhone2023soxs,
        title = {SOXS: simulated observations of X-ray sources},
       author = {ZuHone, John A and Vikhlinin, Alexey and Tremblay, Grant R and Randall, Scott W and Andrade-Santos, Felipe and Bourdin, Herve},
      journal = {Astrophysics Source Code Library},
        pages = {ascl--2301},
         year = {2023},
          url = {https://ui.adsabs.harvard.edu/abs/2023ascl.soft01024Z/abstract}
}

@article{badenes2003thermal,
        title = {Thermal X-ray emission from shocked ejecta in type Ia supernova remnants: prospects for explosion mechanism identification},
       author = {Badenes, Carles and Bravo, Eduardo and Borkowski, Kazimierz J and Dominguez, Inmaculada},
      journal = {The Astrophysical Journal},
       volume = {593},
       number = {1},
        pages = {358},
         year = {2003},
    publisher = {IOP Publishing},
          url = {https://iopscience.iop.org/article/10.1086/376448},
          doi = {10.1086/376448}
}

@article{ferrand2021supernova,
        title = {From supernova to supernova remnant: comparison of thermonuclear explosion models},
       author = {Ferrand, Gilles and Warren, Donald C and Ono, Masaomi and Nagataki, Shigehiro and R{\"o}pke, Friedrich K and Seitenzahl, Ivo R and Lach, Florian and Iwasaki, Hiroyoshi and Sato, Toshiki},
      journal = {The Astrophysical Journal},
       volume = {906},
       number = {2},
        pages = {93},
         year = {2021},
    publisher = {IOP Publishing},
          url = {https://iopscience.iop.org/article/10.3847/1538-4357/abc951},
          doi = {10.3847/1538-4357/abc951}
}

@article{riess1998observational,
        title = {Observational evidence from supernovae for an accelerating universe and a cosmological constant},
       author = {Riess, Adam G and Filippenko, Alexei V and Challis, Peter and Clocchiatti, Alejandro and Diercks, Alan and Garnavich, Peter M and Gilliland, Ron L and Hogan, Craig J and Jha, Saurabh and Kirshner, Robert P and others},
      journal = {The astronomical journal},
       volume = {116},
       number = {3},
        pages = {1009},
         year = {1998},
    publisher = {IOP Publishing},
          url = {https://iopscience.iop.org/article/10.1086/300499},
          doi = {10.1086/300499}
}

@article{perlmutter1999measurements,
        title = {Measurements of $\Omega$ and $\Lambda$ from 42 high-redshift supernovae},
       author = {Perlmutter, Saul and Aldering, Goldhaber and Goldhaber, Gerson and Knop, Richard A and Nugent, Peter and Castro, Patricia G and Deustua, Susana and Fabbro, Sebastien and Goobar, Ariel and Groom, Donald E and others},
      journal = {The Astrophysical Journal},
       volume = {517},
       number = {2},
        pages = {565},
         year = {1999},
    publisher = {IOP Publishing},
          url = {https://iopscience.iop.org/article/10.1086/307221},
          doi = {10.1086/307221}
}

@article{Iben1984Supernovaemass,
        title = {{Supernovae of type I as end products of the evolution of binaries with components of moderate initial mass.}},
       author = {Iben, Icko Jr and Tutukov, Alexander V.},
      journal = {ApJS},
       volume = {54},
        pages = {335--372},
         year = {1984},
          url = {https://ui.adsabs.harvard.edu/abs/1984ApJS...54..335I/abstract},
          doi = {10.1086/190932}
}

@article{webbink1984double,
        title = {Double white dwarfs as progenitors of R Coronae Borealis stars and Type I supernovae},
       author = {Webbink, Ronald F},
      journal = {Astrophysical Journal},
       volume = {277},
       number = {1},
        pages = {355--360},
         year = {1984},
          url = {https://ui.adsabs.harvard.edu/abs/1984ApJ...277..355W/abstract},
          doi = {10.1086/161701}
}

@article{badenes2005thermal,
        title = {Thermal X-ray emission from shocked ejecta in Type Ia supernova remnants. II. Parameters affecting the spectrum},
       author = {Badenes, Carles and Borkowski, Kazimierz J and Bravo, Eduardo},
      journal = {The Astrophysical Journal},
       volume = {624},
       number = {1},
        pages = {198},
         year = {2005},
    publisher = {IOP Publishing},
          url = {https://iopscience.iop.org/article/10.1086/428829},
          doi = {10.1086/428829}
}

@article{badenes2006constraints,
        title = {Constraints on the physics of Type Ia supernovae from the X-ray spectrum of the Tycho supernova remnant},
       author = {Badenes, Carles and Borkowski, Kazimierz J and Hughes, John P and Hwang, Una and Bravo, Eduardo},
       journal = {The Astrophysical Journal},
      volume = {645},
      number = {2},
       pages = {1373},
        year = {2006},
    publisher = {IOP Publishing},
          url = {https://iopscience.iop.org/article/10.1086/504399},
          doi = {10.1086/504399}
}

@article{sato2025missing,
        title = {Missing Titanium in the Asymmetric Supernova Remnant W 49B},
       author = {Sato, Toshiki and Sawada, Makoto and Maeda, Keiichi and Hughes, John P and Williams, Brian J},
      journal = {The Astrophysical Journal},
       volume = {986},
       number = {1},
        pages = {94},
         year = {2025},
    publisher = {IOP Publishing},
          url = {https://iopscience.iop.org/article/10.3847/1538-4357/add932},
          doi = {10.3847/1538-4357/add932}
}

@article{xrism2025kinematic,
          doi = {10.3847/2041-8213/ade138},
          url = {https://doi.org/10.3847/2041-8213/ade138},
         year = {2025},
        month = {jul},
    publisher = {The American Astronomical Society},
       volume = {988},
       number = {2},
        pages = {L58},
       author = {{XRISM Collaboration} and Audard, Marc and Awaki, Hisamitsu and Ballhausen, Ralf and Bamba, Aya and Behar, Ehud and Boissay-Malaquin, Rozenn and Brenneman, Laura and Brown, Gregory V. and Corrales, Lia and Costantini, Elisa and Cumbee, Renata and Díaz Trigo, María and Done, Chris and Dotani, Tadayasu and Ebisawa, Ken and Eckart, Megan E. and Eckert, Dominique and Eguchi, Satoshi and Enoto, Teruaki and Ezoe, Yuichiro and Foster, Adam and Fujimoto, Ryuichi and Fujita, Yutaka and Fukazawa, Yasushi and Fukushima, Kotaro and Furuzawa, Akihiro and Gallo, Luigi and García, Javier A. and Gu, Liyi and Guainazzi, Matteo and Hagino, Kouichi and Hamaguchi, Kenji and Hatsukade, Isamu and Hayashi, Katsuhiro and Hayashi, Takayuki and Hell, Natalie and Hodges-Kluck, Edmund and Hornschemeier, Ann and Ichinohe, Yuto and Ishi, Daiki and Ishida, Manabu and Ishikawa, Kumi and Ishisaki, Yoshitaka and Kaastra, Jelle and Kallman, Timothy and Kara, Erin and Katsuda, Satoru and Kanemaru, Yoshiaki and Kelley, Richard L. and Kilbourne, Caroline A. and Kitamoto, Shunji and Kobayashi, Shogo and Kohmura, Takayoshi and Kubota, Aya and Leutenegger, Maurice A. and Loewenstein, Michael and Maeda, Yoshitomo and Markevitch, Maxim and Matsumoto, Hironori and Matsushita, Kyoko and McCammon, Dan and McNamara, Brian and Mernier, François and Miller, Eric D. and Miller, Jon M. and Mitsuishi, Ikuyuki and Mizumoto, Misaki and Mizuno, Tsunefumi and Mori, Koji and Mukai, Koji and Murakami, Hiroshi and Mushotzky, Richard and Nakajima, Hiroshi and Nakazawa, Kazuhiro and Ness, Jan-Uwe and Nobukawa, Kumiko and Nobukawa, Masayoshi and Noda, Hirofumi and Odaka, Hirokazu and Ogawa, Shoji and Ogorzalek, Anna and Okajima, Takashi and Ota, Naomi and Paltani, Stephane and Petre, Robert and Plucinsky, Paul and Porter, Frederick S. and Pottschmidt, Katja and Sano, Hidetoshi and Sato, Kosuke and Sato, Toshiki and Sawada, Makoto and Seta, Hiromi and Shidatsu, Megumi and Simionescu, Aurora and Smith, Randall and Suzuki, Hiromasa and Szymkowiak, Andrew and Takahashi, Hiromitsu and Takeo, Mai and Tamagawa, Toru and Tamura, Keisuke and Tanaka, Takaaki and Tanimoto, Atsushi and Tashiro, Makoto and Terada, Yukikatsu and Terashima, Yuichi and Tsuboi, Yohko and Tsujimoto, Masahiro and Tsunemi, Hiroshi and Tsuru, Takeshi G. and Tümer, Ayşegül and Uchida, Hiroyuki and Uchida, Nagomi and Uchida, Yuusuke and Uchiyama, Hideki and Ueda, Yoshihiro and Uno, Shinichiro and Vink, Jacco and Watanabe, Shin and Williams, Brian J. and Yamada, Satoshi and Yamada, Shinya and Yamaguchi, Hiroya and Yamaoka, Kazutaka and Yamasaki, Noriko and Yamauchi, Makoto and Yamauchi, Shigeo and Yaqoob, Tahir and Yoneyama, Tomokage and Yoshida, Tessei and Yukita, Mihoko and Zhuravleva, Irina and Amano, Yuki and Gall, Amy and Mitrani, Sharon and Murakami, Kaito and Rahin, Roi and Suzuki, Nari},
        title = {Kinematic Evidence for Bipolar Ejecta Flows in the Galactic Supernova Remnant W49B},
      journal = {The Astrophysical Journal Letters}
}

@article{Court_2024,
          doi = {10.3847/1538-4357/ad165f},
          url = {https://doi.org/10.3847/1538-4357/ad165f},
         year = {2024},
        month = {feb},
    publisher = {The American Astronomical Society},
       volume = {962},
       number = {1},
        pages = {63},
       author = {Court, Travis and Badenes, Carles and Lee, Shiu-Hang and Patnaude, Daniel and García-Segura, Guillermo and Bravo, Eduardo},
        title = {Do Type Ia Supernovae Explode inside Planetary Nebulae?},
      journal = {The Astrophysical Journal}
}

@article{court2026classification,
doi = {10.3847/1538-4357/ae1a7e},
url = {https://doi.org/10.3847/1538-4357/ae1a7e},
year = {2025},
month = {dec},
publisher = {The American Astronomical Society},
volume = {996},
number = {1},
pages = {100},
author = {Court, Travis and Badenes, Carles and Lee, Shiu-Hang and Patnaude, Daniel and Bravo, Eduardo},
title = {A Classification Scheme for X-Ray-bright Type Ia Supernova Remnants Based on Their Circumstellar Interaction},
journal = {The Astrophysical Journal}
}

@article{whelan1973binaries,
        title = {Binaries and supernovae of type I},
       author = {Whelan, John and Iben, Icko},
      journal = {Astrophysical Journal, Vol. 186, pp. 1007-1014 (1973)},
       volume = {186},
        pages = {1007--1014},
         year = {1973},
          url = {https://ui.adsabs.harvard.edu/abs/1973ApJ...186.1007W/abstract},
          doi = {10.1086/152565}
}

@article{nomoto1982accreting,
        title = {Accreting white dwarf models for type I supernovae. I-Presupernova evolution and triggering mechanisms},
       author = {Nomoto, K},
      journal = {Astrophysical Journal, Part 1, vol. 253, Feb. 15, 1982, p. 798-810.},
       volume = {253},
        pages = {798--810},
         year = {1982},
          url = {https://ui.adsabs.harvard.edu/abs/1982ApJ...253..798N/abstract},
          doi = {10.1086/159682}
}

@article{ferrand2025role,
          doi = {10.3847/1538-4357/ae1602},
          url = {https://doi.org/10.3847/1538-4357/ae1602},
         year = {2025},
        month = {dec},
    publisher = {The American Astronomical Society},
       volume = {995},
       number = {1},
        pages = {85},
       author = {Ferrand, Gilles and Pakmor, Rüdiger and Fujimaru, Yusei and Lee, Shiu-Hang and Safi-Harb, Samar and Nagataki, Shigehiro and Röpke, Friedrich K. and Decourchelle, Anne and Seitenzahl, Ivo R. and Patnaude, Daniel},
        title = {The Role of the Secondary White Dwarf in a Double-degenerate Double-detonation Explosion, in the Supernova Remnant Phase},
      journal = {The Astrophysical Journal}
}

@article{yamaguchi2021discovery,
          doi = {10.3847/2041-8213/abee8a},
          url = {https://doi.org/10.3847/2041-8213/abee8a},
         year = {2021},
        month = {apr},
    publisher = {The American Astronomical Society},
       volume = {910},
       number = {2},
        pages = {L24},
       author = {Yamaguchi, Hiroya and Acero, Fabio and Li, Chuan-Jui and Chu, You-Hua},
        title = {Discovery of Double-ring Structure in the Supernova Remnant N103B: Evidence for Bipolar Winds from a Type Ia Supernova Progenitor},
      journal = {The Astrophysical Journal Letters}
}

@article{safi2005chandra,
          doi = {10.1086/425960},
          url = {https://doi.org/10.1086/425960},
         year = {2005},
        month = {jan},
    publisher = {},
       volume = {618},
       number = {1},
        pages = {321},
       author = {Safi-Harb, S. and Dubner, G. and Petre, R. and Holt, S. S. and Durouchoux, P.},
        title = {Chandra Spatially Resolved Spectroscopic Study and Multiwavelength Imaging of the Supernova Remnant 3C 397 (G41.1–0.3)},
      journal = {The Astrophysical Journal}
}

@article{martinez2020evidence,
        title = {Evidence of a Type Ia Progenitor for Supernova Remnant 3C 397},
       author = {Mart{\'\i}nez-Rodr{\'\i}guez, H{\'e}ctor and Lopez, Laura A and Auchettl, Katie and Badenes, Carles and Holland-Ashford, Tyler and Patnaude, Daniel J and Lee, Shiu-Hang and Foster, Adam R and Slane, Patrick O},
      journal = {arXiv preprint arXiv:2006.08681},
         year = {2020},
          doi = {10.48550/arXiv.2006.08681}
}

@article{guest2023rapid,
          doi = {10.3847/1538-4357/acbf4e},
          url = {https://doi.org/10.3847/1538-4357/acbf4e},
         year = {2023},
        month = {mar},
    publisher = {The American Astronomical Society},
       volume = {946},
       number = {1},
        pages = {44},
       author = {Guest, Benson T. and Borkowski, Kazimierz J. and Ghavamian, Parviz and Petre, Robert and Picquenot, Adrien and Reynolds, Stephen P. and Seitenzahl, Ivo R. and Williams, Brian J.},
        title = {Rapid Expansion of the Young Type Ia Supernova Remnant 0519–69.0: More Evidence for a Circumstellar Shell},
      journal = {The Astrophysical Journal}
}

@article{zhou2018asymmetric,
       author = {Zhou, Ping and Vink, Jacco},
    	title = {Asymmetric Type-Ia supernova origin of W49B as revealed from spatially resolved X-ray spectroscopic study},
    	   doi = {10.1051/0004-6361/201731583},
    	   url = {https://doi.org/10.1051/0004-6361/201731583},
      journal = {A\&A},
         year = 2018,
       volume = 615,
    	pages = "A150",
}

@article{raymond2023electron,
          doi = {10.3847/1538-4357/acc528},
          url = {https://doi.org/10.3847/1538-4357/acc528},
         year = {2023},
        month = {may},
    publisher = {The American Astronomical Society},
       volume = {949},
       number = {2},
        pages = {50},
       author = {Raymond, John C. and Ghavamian, Parviz and Bohdan, Artem and Ryu, Dongsu and Niemiec, Jacek and Sironi, Lorenzo and Tran, Aaron and Amato, Elena and Hoshino, Masahiro and Pohl, Martin and Amano, Takanobu and Fiuza, Frederico},
        title = {Electron–Ion Temperature Ratio in Astrophysical Shocks},
      journal = {The Astrophysical Journal}
}

@article{ferrand20103d,
        title = {3D simulations of supernova remnants evolution including non-linear particle acceleration},
       author = {Ferrand, Gilles and Decourchelle, Anne and Ballet, Jean and Teyssier, Romain and Fraschetti, Federico},
      journal = {Astronomy \& Astrophysics},
       volume = {509},
        pages = {L10},
         year = {2010},
    publisher = {EDP Sciences},
          url = {https://www.aanda.org/articles/aa/abs/2010/01/aa13666-09/aa13666-09.html},
          doi = {10.1051/0004-6361/200913666}
}

@article{ruiter2025type,
        title = {Type Ia supernova progenitors: a contemporary view of a long-standing puzzle},
       author = {Ruiter, Ashley Jade and Seitenzahl, Ivo Rolf},
      journal = {The Astronomy and Astrophysics Review},
       volume = {33},
       number = {1},
        pages = {1--97},
         year = {2025},
    publisher = {Springer},
          url = {https://link.springer.com/article/10.1007/s00159-024-00158-9},
          doi = {10.1007/s00159-024-00158-9}
}

@article{smith2005astrophysical,
        title = {The Astrophysical Plasma Emission Database: Progress and Plans},
       author = {Smith, Randall K and Brickhouse, Nancy S and Liedahl, Duane A},
      journal = {Highlights of Astronomy},
       volume = {13},
        pages = {666--667},
         year = {2005},
    publisher = {Cambridge University Press},
          url = {https://www.cambridge.org/core/journals/highlights-of-astronomy/article/astrophysical-plasma-emission-database-progress-and-plans/560CB5A3828BC949062868AB07F253CD},
          doi = {10.1017/S1539299600016786}
}

@article{cruise2025newathena,
        title = {The NewAthena mission concept in the context of the next decade of X-ray astronomy},
       author = {Cruise, Mike and Guainazzi, Matteo and Aird, James and Carrera, Francisco J and Costantini, Elisa and Corrales, Lia and Dauser, Thomas and Eckert, Dominique and Gastaldello, Fabio and Matsumoto, Hironori and others},
      journal = {Nature Astronomy},
       volume = {9},
       number = {1},
        pages = {36--44},
         year = {2025},
    publisher = {Nature Publishing Group UK London},
          url = {https://www.nature.com/articles/s41550-024-02416-3},
          doi = {10.1038/s41550-024-02416-3}
}

@article{garcia2018common,
        title = {Common envelope shaping of planetary nebulae},
       author = {Garc{\'\i}a-Segura, Guillermo and Ricker, Paul M and Taam, Ronald E},
      journal = {The Astrophysical Journal},
       volume = {860},
       number = {1},
        pages = {19},
         year = {2018},
    publisher = {IOP Publishing},
          url = {https://iopscience.iop.org/article/10.3847/1538-4357/aac08c},
          doi = {10.3847/1538-4357/aac08c}
}

@article{yamaguchi2013new,
          doi = {10.1088/0004-637X/780/2/136},
          url = {https://doi.org/10.1088/0004-637X/780/2/136},
         year = {2013},
        month = {dec},
    publisher = {The American Astronomical Society},
       volume = {780},
       number = {2},
        pages = {136},
       author = {Yamaguchi, Hiroya and Eriksen, Kristoffer A. and Badenes, Carles and Hughes, John P. and Brickhouse, Nancy S. and Foster, Adam R. and Patnaude, Daniel J. and Petre, Robert and Slane, Patrick O. and Smith, Randall K.},
        title = {NEW EVIDENCE FOR EFFICIENT COLLISIONLESS HEATING OF ELECTRONS AT THE REVERSE SHOCK OF A YOUNG SUPERNOVA REMNANT},
      journal = {The Astrophysical Journal}
}

@article{fujimaru2026probing,
          doi = {10.3847/1538-4357/ae4004},
          url = {https://doi.org/10.3847/1538-4357/ae4004},
         year = {2026},
        month = {mar},
    publisher = {The American Astronomical Society},
       volume = {999},
       number = {2},
        pages = {151},
       author = {Fujimaru, Yusei and Lee, Shiu-Hang and Ferrand, Gilles and Patnaude, Daniel and Nagataki, Shigehiro and Pakmor, Rüdiger and Safi-Harb, Samar and Röpke, Friedrich K. and Decourchelle, Anne and Seitenzahl, Ivo R.},
        title = {Probing the Diversity of Type Ia Supernova Remnants in 3D Hydrodynamic Simulations with X-Ray Spectral Synthesis},
      journal = {The Astrophysical Journal}
}

@article{kawashima2026single,
        title = {Single vs. Binary Origin: The Diversity of Stripped-Envelope Supernova Remnants},
       author = {Kawashima, Gaku and Lee, Shiu-Hang and Maeda, Keiichi and Patnaude, Daniel},
      journal = {arXiv preprint arXiv:2603.28234},
         year = {2026}
}

@article{katsuragawa2022formation,
          doi = {10.3847/1538-4357/ac8cf9},
          url = {https://doi.org/10.3847/1538-4357/ac8cf9},
         year = {2022},
        month = {oct},
    publisher = {The American Astronomical Society},
       volume = {938},
       number = {1},
        pages = {23},
       author = {Katsuragawa, Miho and Lee, Shiu-Hang and Odaka, Hirokazu and Bamba, Aya and Matsumura, Hideaki and Takahashi, Tadayuki},
        title = {On the Formation of Over-ionized Plasma in Evolved Supernova Remnants},
      journal = {The Astrophysical Journal}
}

@article{keohane2007near,
          doi = {10.1086/509311},
          url = {https://doi.org/10.1086/509311},
         year = {2007},
        month = {jan},
    publisher = {},
       volume = {654},
       number = {2},
        pages = {938},
       author = {Keohane, Jonathan W. and Reach, William T. and Rho, Jeonghee and Jarrett, Thomas H.},
        title = {A Near-Infrared and X-Ray Study of W49 B:  A Wind Cavity Explosion},
      journal = {The Astrophysical Journal}
}

@article{miceli2010spatial,
       author = {Miceli, M. and Bocchino, F. and Decourchelle, A. and Ballet, J. and Reale, F.},
        title = {Spatial identification of the overionized plasma in W49B},
           DOI= {10.1051/0004-6361/200913713},
           url= {https://doi.org/10.1051/0004-6361/200913713},
      journal = {A\&A},
         year = {2010},
       volume = {514},
        pages = {L2},
        month = {may},
}

@article{lopez2013unraveling,
          doi = {10.1088/0004-637X/777/2/145},
          url = {https://doi.org/10.1088/0004-637X/777/2/145},
         year = {2013},
        month = {oct},
    publisher = {The American Astronomical Society},
       volume = {777},
       number = {2},
        pages = {145},
       author = {Lopez, Laura A. and Pearson, Sarah and Ramirez-Ruiz, Enrico and Castro, Daniel and Yamaguchi, Hiroya and Slane, Patrick O. and Smith, Randall K.},
        title = {UNRAVELING THE ORIGIN OF OVERIONIZED PLASMA IN THE GALACTIC SUPERNOVA REMNANT W49B},
      journal = {The Astrophysical Journal}
}

@article{yamaguchi2018evidence,
          doi = {10.3847/2041-8213/aaf055},
          url = {https://doi.org/10.3847/2041-8213/aaf055},
         year = {2018},
        month = {nov},
    publisher = {The American Astronomical Society},
       volume = {868},
       number = {2},
        pages = {L35},
       author = {Yamaguchi, Hiroya and Tanaka, Takaaki and Wik, Daniel R. and Rho, Jeonghee and Bamba, Aya and Castro, Daniel and Smith, Randall K. and Foster, Adam R. and Uchida, Hiroyuki and Petre, Robert and Williams, Brian J.},
        title = {Evidence for Rapid Adiabatic Cooling as an Origin of the Recombining Plasma in the Supernova Remnant W49B Revealed by NuSTAR Observations},
      journal = {The Astrophysical Journal Letters}
}

@article{ozawa2009suzaku,
          doi = {10.1088/0004-637X/706/1/L71},
          url = {https://doi.org/10.1088/0004-637X/706/1/L71},
         year = {2009},
        month = {oct},
    publisher = {The American Astronomical Society},
       volume = {706},
       number = {1},
        pages = {L71},
       author = {Ozawa, M. and Koyama, K. and Yamaguchi, H. and Masai, K. and Tamagawa, T.},
        title = {SUZAKU DISCOVERY OF THE STRONG RADIATIVE RECOMBINATION CONTINUUM OF IRON FROM THE SUPERNOVA REMNANT W49B},
      journal = {The Astrophysical Journal}
}

@article{holland2026xrism,
          doi = {10.3847/1538-4357/ae4357},
          url = {https://doi.org/10.3847/1538-4357/ae4357},
         year = {2026},
        month = {mar},
    publisher = {The American Astronomical Society},
       volume = {999},
       number = {2},
        pages = {205},
       author = {Holland-Ashford, Tyler and Williams, Brian J.},
        title = {XRISM Observation of the Supernova Remnant N103B: Velocity Structure and Thermal Properties},
      journal = {The Astrophysical Journal}
}

@article{liu2023type,
          doi = {10.1088/1674-4527/acd89e},
          url = {https://doi.org/10.1088/1674-4527/acd89e},
         year = {2023},
        month = {jul},
    publisher = {National Astromonical Observatories, CAS and IOP Publishing},
       volume = {23},
       number = {8},
        pages = {082001},
       author = {Liu, Zheng-Wei and Röpke, Friedrich K. and Han, Zhanwen},
        title = {Type Ia Supernova Explosions in Binary Systems: A Review},
      journal = {Research in Astronomy and Astrophysics}
}

@article{dilday2012ptf,
       author = {B. Dilday  and D. A. Howell  and S. B. Cenko  and J. M. Silverman  and P. E. Nugent  and M. Sullivan  and S. Ben-Ami  and L. Bildsten  and M. Bolte  and M. Endl  and A. V. Filippenko  and O. Gnat  and A. Horesh  and E. Hsiao  and M. M. Kasliwal  and D. Kirkman  and K. Maguire  and G. W. Marcy  and K. Moore  and Y. Pan  and J. T. Parrent  and P. Podsiadlowski  and R. M. Quimby  and A. Sternberg  and N. Suzuki  and D. R. Tytler  and D. Xu  and J. S. Bloom  and A. Gal-Yam  and I. M. Hook  and S. R. Kulkarni  and N. M. Law  and E. O. Ofek  and D. Polishook  and D. Poznanski },
        title = {PTF 11kx: A Type Ia Supernova with a Symbiotic Nova Progenitor},
      journal = {Science},
       volume = {337},
       number = {6097},
        pages = {942-945},
         year = {2012},
          doi = {10.1126/science.1219164},
          URL = {https://www.science.org/doi/abs/10.1126/science.1219164},
       eprint = {https://www.science.org/doi/pdf/10.1126/science.1219164}
}

@article{silverman2013type,
          doi = {10.1088/0067-0049/207/1/3},
          url = {https://doi.org/10.1088/0067-0049/207/1/3},
         year = {2013},
        month = {jun},
    publisher = {The American Astronomical Society},
       volume = {207},
       number = {1},
        pages = {3},
       author = {Silverman, Jeffrey M. and Nugent, Peter E. and Gal-Yam, Avishay and Sullivan, Mark and Howell, D. Andrew and Filippenko, Alexei V. and Arcavi, Iair and Ben-Ami, Sagi and Bloom, Joshua S. and Cenko, S. Bradley and Cao, Yi and Chornock, Ryan and Clubb, Kelsey I. and Coil, Alison L. and Foley, Ryan J. and Graham, Melissa L. and Griffith, Christopher V. and Horesh, Assaf and Kasliwal, Mansi M. and Kulkarni, Shrinivas R. and Leonard, Douglas C. and Li, Weidong and Matheson, Thomas and Miller, Adam A. and Modjaz, Maryam and Ofek, Eran O. and Pan, Yen-Chen and Perley, Daniel A. and Poznanski, Dovi and Quimby, Robert M. and Steele, Thea N. and Sternberg, Assaf and Xu, Dong and Yaron, Ofer},
        title = {TYPE Ia SUPERNOVAE STRONGLY INTERACTING WITH THEIR CIRCUMSTELLAR MEDIUM},
      journal = {The Astrophysical Journal Supplement Series}
}

@article{sharma2023systematic,
          doi = {10.3847/1538-4357/acbc16},
          url = {https://doi.org/10.3847/1538-4357/acbc16},
         year = {2023},
        month = {may},
    publisher = {The American Astronomical Society},
       volume = {948},
       number = {1},
        pages = {52},
       author = {Sharma, Yashvi and Sollerman, Jesper and Fremling, Christoffer and Kulkarni, Shrinivas R. and De, Kishalay and Irani, Ido and Schulze, Steve and Strotjohann, Nora Linn and Gal-Yam, Avishay and Maguire, Kate and Perley, Daniel A. and Bellm, Eric C. and Kool, Erik C. and Brink, Thomas G. and Bruch, Rachel and Deckers, Maxime and Dekany, Richard and Dugas, Alison and Filippenko, Alexei V. and Goldwasser, Samantha and Graham, Matthew J. and Graham, Melissa L. and Groom, Steven L. and Hankins, Matt and Jencson, Jacob and Johansson, Joel P. and Karambelkar, Viraj and Kasliwal, Mansi M. and Masci, Frank J. and Medford, Michael S. and Neill, James D. and Nir, Guy and Riddle, Reed L. and Rigault, Mickael and Schweyer, Tassilo and Terwel, Jacco H. and Yan, Lin and Yang, Yi and Yao, Yuhan},
        title = {A Systematic Study of Ia-CSM Supernovae from the ZTF Bright Transient Survey},
      journal = {The Astrophysical Journal}
}

@article{soker2026type,
	author={Soker, Noam},
	title={Type Ia supernovae interacting with a close circumstellar material (SNe Ia-CSM) are SNe Ia inside planetary nebulae (SNIPs)},
	journal={Research in Astronomy and Astrophysics},
    doi={10.1088/1674-4527/ae77c1},
	url={http://iopscience.iop.org/article/10.1088/1674-4527/ae77c1},
	year={2026}
}

@article{tsebrenko2015modelling,
       author = {Tsebrenko, Danny and Soker, Noam},
          doi = {10.1093/mnras/stv669},
       eprint = {https://academic.oup.com/mnras/article-pdf/450/2/1399/3033967/stv669.pdf},
         issn = {0035-8711},
      journal = {Monthly Notices of the Royal Astronomical Society},
        month = {06},
       number = {2},
        pages = {1399-1408},
        title = {Modelling SNR G1.9+0.3 as a Supernova inside a Planetary Nebula},
          url = {https://doi.org/10.1093/mnras/stv669},
       volume = {450},
         year = {2015}
}

@article{tsebrenko2013type,
author = {Tsebrenko, Danny and Soker, Noam},
          doi = {10.1093/mnras/stt1301},
eprint = {https://academic.oup.com/mnras/article-pdf/435/1/320/3876428/stt1301.pdf},
issn = {0035-8711},
journal = {Monthly Notices of the Royal Astronomical Society},
month = {10},
number = {1},
pages = {320-328},
title = {Type Ia supernovae inside planetary nebulae: shaping by jets},
url = {https://doi.org/10.1093/mnras/stt1301},
volume = {435},
year = {2013}
}

@article{martinez2018chandrasekhar,
  title = {Chandrasekhar and Sub-Chandrasekhar Models for the x-Ray Emission of Type Ia Supernova Remnants. {{I}}. {{Bulk}} Properties},
  author = {{Mart{\'i}nez-Rodr{\'i}guez}, H{\'e}ctor and Badenes, Carles and Lee, Shiu-Hang and Patnaude, Daniel J. and Foster, Adam R. and Yamaguchi, Hiroya and Auchettl, Katie and Bravo, Eduardo and Slane, Patrick O. and Piro, Anthony L. and Park, Sangwook and Nagataki, Shigehiro},
  year = 2018,
  month = oct,
  journal = {The Astrophysical Journal},
  volume = {865},
  number = {2},
  pages = {151},
  publisher = {The American Astronomical Society},
  doi = {10.3847/1538-4357/aadaec},
  url = {https://doi.org/10.3847/1538-4357/aadaec}
}

@article{kashi2011circumbinary,
title = {A circumbinary disc in the final stages of common envelope and the core-degenerate scenario for {Type} {Ia} supernovae},
volume = {417},
issn = {0035-8711},
url = {https://doi.org/10.1111/j.1365-2966.2011.19361.x},
doi = {10.1111/j.1365-2966.2011.19361.x},
number = {2},
journal = {Monthly Notices of the Royal Astronomical Society},
author = {Kashi, Amit and Soker, Noam},
month = oct,
year = {2011},
note = {\_eprint: https://academic.oup.com/mnras/article-pdf/417/2/1466/3074873/mnras0417-1466.pdf},
pages = {1466--1479},
}

@article{itoh1989effect,
  title = {The Effect of a Circumstellar Medium on the {{X-ray}} Emission of Young Remnants of {{Type II}} Supernovae},
  author = {Itoh, Hiroshi and Masai, Kuniaki},
  year = 1989,
  month = feb,
  journal = {Monthly Notices of the Royal Astronomical Society},
  volume = {236},
  number = {4},
  eprint = {https://academic.oup.com/mnras/article-pdf/236/4/885/3924818/mnras236-0885.pdf},
  pages = {885--899},
  issn = {0035-8711},
  doi = {10.1093/mnras/236.4.885},
  url = {https://doi.org/10.1093/mnras/236.4.885}
}

@article{badenes2007are,
  title = {Are the Models for Type Ia Supernova Progenitors Consistent with the Properties of Supernova Remnants?},
  author = {Badenes, Carles and Hughes, John P. and Bravo, Eduardo and Langer, Norbert},
  year = 2007,
  month = jun,
  journal = {The Astrophysical Journal},
  volume = {662},
  number = {1},
  pages = {472},
  doi = {10.1086/518022},
  url = {https://doi.org/10.1086/518022}
}

@article{jerkstrand2020type,
  title = {A Type {{Ia}} Supernova at the Heart of Superluminous Transient {{SN}} 2006gy},
  author = {Jerkstrand, Anders and Maeda, Keiichi and Kawabata, Koji S.},
  year = 2020,
  journal = {Science},
  volume = {367},
  number = {6476},
  eprint = {https://www.science.org/doi/pdf/10.1126/science.aaw1469},
  pages = {415--418},
  doi = {10.1126/science.aaw1469},
  url = {https://www.science.org/doi/abs/10.1126/science.aaw1469}
}

@article{uno2023sn,
  title = {{{SN}} 2020uem: A Possible Thermonuclear Explosion within a Dense Circumstellar Medium. {{I}}. {{The}} Nature of Type Iin/Ia-{{CSM}} Sne from Photometry and Spectroscopy},
  author = {Uno, Kohki and Maeda, Keiichi and Nagao, Takashi and Nakaoka, Tatsuya and Motohara, Kentaro and Tajitsu, Akito and Konishi, Masahiro and Koyama, Shuhei and Takahashi, Hidenori and Tanaka, Masaomi and Kuncarayakti, Hanindyo and Kawabata, Miho and Yamanaka, Masayuki and Aoki, Kentaro and Isogai, Keisuke and Taguchi, Kenta and Ogawa, Mao and Kawabata, Koji S. and Yoshii, Yuzuru and Miyata, Takashi and Imazawa, Ryo},
  year = 2023,
  month = feb,
  journal = {The Astrophysical Journal},
  volume = {944},
  number = {2},
  pages = {203},
  publisher = {The American Astronomical Society},
  doi = {10.3847/1538-4357/acb5ec},
  url = {https://doi.org/10.3847/1538-4357/acb5ec}
}

@article{ghavamian2013electronion,
  title = {Electron-{{Ion Temperature Equilibration}} in {{Collisionless Shocks}}: {{The Supernova Remnant-Solar Wind Connection}}},
  author = {Ghavamian, Parviz and Schwartz, Steven J. and Mitchell, Jeremy and Masters, Adam and Laming, J. Martin},
  year = 2013,
  month = oct,
  journal = {Space Science Reviews},
  volume = {178},
  number = {2},
  pages = {633--663},
  issn = {1572-9672},
  doi = {10.1007/s11214-013-9999-0},
  url = {https://doi.org/10.1007/s11214-013-9999-0}
}

@article{ricker2012amr,
  title = {An Amr Study of the Common-Envelope Phase of Binary Evolution},
  author = {Ricker, Paul M. and Taam, Ronald E.},
  year = 2012,
  month = jan,
  journal = {The Astrophysical Journal},
  volume = {746},
  number = {1},
  pages = {74},
  publisher = {The American Astronomical Society},
  doi = {10.1088/0004-637X/746/1/74},
  url = {https://doi.org/10.1088/0004-637X/746/1/74}
}

@article{dimitriadis2025ztf,
  title = {{{ZTF SN Ia DR2}}: {{The}} Diversity and Relative Rates of the Thermonuclear Supernova Population},
  author = {Dimitriadis, G. and Burgaz, U. and Deckers, M. and Maguire, K. and Johansson, J. and Smith, M. and Rigault, M. and Frohmaier, C. and Sollerman, J. and Galbany, L. and Kim, Y.-L. and Liu, C. and Miller, A. A. and Nugent, P. E. and Alburai, A. and Chen, P. and Dhawan, S. and Ginolin, M. and Goobar, A. and Groom, S. L. and Harvey, L. and Kenworthy, W. D. and Kulkarni, S. R. and Phan, K. and Popovic, B. and Riddle, R. L. and Rusholme, B. and M\"uller-Bravo, T. E. and Nordin, J. and Terwel, J. H. and Townsend, A.},
  year = 2025,
  journal = {Astronomy \& Astrophysics},
  volume = {694},
  pages = {A10},
  doi = {10.1051/0004-6361/202451852},
  url = {https://doi.org/10.1051/0004-6361/202451852}
}

@article{xrism2025overionized,
  title = {Overionized Plasma in the Supernova Remnant {{Sagittarius A East}} Anchored by {{XRISM}} Observations},
  author = {{XRISM Collaboration} and Audard, Marc and Awaki, Hisamitsu and Ballhausen, Ralf and Bamba, Aya and Behar, Ehud and {Boissay-Malaquin}, Rozenn and Brenneman, Laura and Brown, Gregory V and Corrales, Lia and Costantini, Elisa and Cumbee, Renata and {Diaz-Trigo}, Maria and Done, Chris and Dotani, Tadayasu and Ebisawa, Ken and Eckart, Megan and Eckert, Dominique and Enoto, Teruaki and Eguchi, Satoshi and Ezoe, Yuichiro and Foster, Adam and Fujimoto, Ryuichi and Fujita, Yutaka and Fukazawa, Yasushi and Fukushima, Kotaro and Furuzawa, Akihiro and Gallo, Luigi and Garc{\'i}a, Javier A and Gu, Liyi and Guainazzi, Matteo and Hagino, Kouichi and Hamaguchi, Kenji and Hatsukade, Isamu and Hayashi, Katsuhiro and Hayashi, Takayuki and Hell, Natalie and {Hodges-Kluck}, Edmund and Hornschemeier, Ann and Ichinohe, Yuto and Ishida, Manabu and Ishikawa, Kumi and Ishisaki, Yoshitaka and Kaastra, Jelle and Kallman, Timothy and Kara, Erin and Katsuda, Satoru and Kanemaru, Yoshiaki and Kelley, Richard and Kilbourne, Caroline and Kitamoto, Shunji and Kobayashi, Shogo and Kohmura, Takayoshi and Kubota, Aya and Leutenegger, Maurice and Loewenstein, Michael and Maeda, Yoshitomo and Markevitch, Maxim and Matsumoto, Hironori and Matsushita, Kyoko and McCammon, Dan and McNamara, Brian and Mernier, Fran{\c c}ois and Miller, Eric D and Miller, Jon M and Mitsuishi, Ikuyuki and Mizumoto, Misaki and Mizuno, Tsunefumi and Mori, Koji and Mukai, Koji and Murakami, Hiroshi and Mushotzky, Richard and Nakajima, Hiroshi and Nakazawa, Kazuhiro and Ness, Jan-Uwe and Nobukawa, Kumiko and Nobukawa, Masayoshi and Noda, Hirofumi and Odaka, Hirokazu and Ogawa, Shoji and Ogorzalek, Anna and Okajima, Takashi and Ota, Naomi and Paltani, Stephane and Petre, Robert and Plucinsky, Paul and Scott Porter, Frederick and Pottschmidt, Katja and Sato, Kosuke and Sato, Toshiki and Sawada, Makoto and Seta, Hiromi and Shidatsu, Megumi and Simionescu, Aurora and Smith, Randall and Suzuki, Hiromasa and Szymkowiak, Andrew and Takahashi, Hiromitsu and Takeo, Mai and Tamagawa, Toru and Tamura, Keisuke and Tanaka, Takaaki and Tanimoto, Atsushi and Tashiro, Makoto and Terada, Yukikatsu and Terashima, Yuichi and Tsuboi, Yohko and Tsujimoto, Masahiro and Tsunemi, Hiroshi and Tsuru, Takeshi G and Uchida, Hiroyuki and Uchida, Nagomi and Uchida, Yuusuke and Uchiyama, Hideki and Ueda, Yoshihiro and Uno, Shinichiro and Vink, Jacco and Watanabe, Shin and Williams, Brian J and Yamada, Satoshi and Yamada, Shinya and Yamaguchi, Hiroya and Yamaoka, Kazutaka and Yamasaki, Noriko and Yamauchi, Makoto and Yamauchi, Shigeo and Yaqoob, Tahir and Yoneyama, Tomokage and Yoshida, Tessei and Yukita, Mihoko and Zhuravleva, Irina and Wang, Q Daniel and Amano, Yuki and Tanaka, Kojiro and Narita, Takuto and Ohshiro, Yuken and Yoshimoto, Anje and Aoki, Yuma and Balakrishnan, Mayura},
  year = 2025,
  month = feb,
  journal = {Publications of the Astronomical Society of Japan},
  volume = {77},
  number = {1},
  eprint = {https://academic.oup.com/pasj/article-pdf/77/1/L1/61276638/psae111.pdf},
  pages = {L1-L8},
  issn = {2053-051X},
  doi = {10.1093/pasj/psae111},
  url = {https://doi.org/10.1093/pasj/psae111}
}

@article{ondratschek2022bipolar,
  title = {Bipolar Planetary Nebulae from Common-Envelope Evolution of Binary Stars{$\star$}},
  author = {Ondratschek, Patrick A. and R\"opke, Friedrich K. and Schneider, Fabian R. N. and Fendt, Christian and Sand, Christian and Ohlmann, Sebastian T. and Pakmor, R\"udiger and Springel, Volker},
  year = 2022,
  journal = {Astronomy \& Astrophysics},
  volume = {660},
  pages = {L8},
  doi = {10.1051/0004-6361/202142478},
  url = {https://doi.org/10.1051/0004-6361/202142478}
}

@article{kozyreva2024thermonuclear,
  title = {Thermonuclear Explosions as Type {{II}} Supernovae{$\star$}},
  author = {Kozyreva, Alexandra and Mor\'an-Fraile, Javier and Holas, Alexander and Bronner, Vincent A. and R\"opke, Friedrich K. and Pavlyuk, Nikolay and Mironov, Alexey and Tsvetkov, Dmitry},
  year = 2024,
  journal = {Astronomy \& Astrophysics},
  volume = {684},
  pages = {A97},
  doi = {10.1051/0004-6361/202348281},
  url = {https://doi.org/10.1051/0004-6361/202348281}
}

@article{Soker2024Supernovae,
	author = {Soker, Noam},
	journal = {The Open Journal of Astrophysics},
	doi = {10.33232/001c.117147},
	year = {2024},
	month = {apr 29},
	publisher = {Maynooth Academic Publishing},
	title = {Supernovae in 2023 (review): possible breakthroughs by late observations},
	volume = {7},
}

@misc{desai2026supernova,
  title = {Supernova Rates and Luminosity Functions from {{ASAS-SN III}}: {{Over}} a Decade of Type Ia Sne and Their Subtypes},
  author = {Desai, Dhvanil D. and Shappee, Benjamin J. and Kochanek, Christopher S. and Stanek, Krzysztof Z. and Ashall, Chris and Beacom, John F. and Burns, Christopher R. and Do, Aaron and Dong, Subo and Hoogendam, Willem B. and Lu, Jing and Pessi, Thallis and Prieto, Jose L. and Thompson, Todd A.},
  year = 2026,
  eprint = {2602.00223},
  primaryclass = {astro-ph.HE},
  url = {https://arxiv.org/abs/2602.00223},
  archiveprefix = {arXiv}
}

@misc{borges2026stars,
  title = {Stars with Plumbing Issues: {{The}} Formation of Collimated Outflows on Common-Envelope Simulations and Comparison to Water Fountains Observations},
  author = {Borges, Sarah V. and Chang, Philip},
  year = 2026,
  eprint = {2603.16667},
  primaryclass = {astro-ph.SR},
  url = {https://arxiv.org/abs/2603.16667},
  archiveprefix = {arXiv}
}

@article{chiotellis2020interaction,
  title = {The Interaction of Type Ia Supernovae with Planetary Nebulae: {{The}} Case of Kepler's Supernova Remnant},
  author = {Chiotellis, A. and Boumis, P. and Spetsieri, Z. T.},
  year = 2020,
  journal = {Galaxies},
  volume = {8},
  number = {38},
  issn = {2075-4434},
  doi = {10.3390/galaxies8020038},
  url = {https://www.mdpi.com/2075-4434/8/2/38}
}

@article{soker2019common,
  title = {Common Envelope to Explosion Delay Time of {{Type Ia}} Supernovae},
  author = {Soker, Noam},
  year = 2019,
  month = dec,
  journal = {Monthly Notices of the Royal Astronomical Society},
  volume = {490},
  number = {2},
  eprint = {https://academic.oup.com/mnras/article-pdf/490/2/2430/30257826/stz2817.pdf},
  pages = {2430--2435},
  issn = {0035-8711},
  doi = {10.1093/mnras/stz2817},
  url = {https://doi.org/10.1093/mnras/stz2817}
}

@article{soker2022common,
  title = {Common Envelope to Explosion Delay Time Distribution ({{CEEDTD}}) of Type Ia Supernovae},
  author = {Soker, Noam},
  year = 2022,
  month = feb,
  journal = {Research in Astronomy and Astrophysics},
  volume = {22},
  number = {3},
  pages = {035025},
  publisher = {{National Astromonical Observatories, CAS and IOP Publishing}},
  doi = {10.1088/1674-4527/ac4d25},
  url = {https://doi.org/10.1088/1674-4527/ac4d25}
}

@article{schindelheim2024snr,
  title = {{{SNR}}\,0519-69.0 as a Type Ia Explosion in a Planetary Nebula Cocoon},
  author = {Schindelheim, Peri and Court, Travis and Badenes, Carles and Lee, Shiu-Hang and Patnaude, Daniel and {Garc{\'i}a-Segura}, Guillermo and Bravo, Eduardo},
  year = 2024,
  month = dec,
  journal = {Research Notes of the AAS},
  volume = {8},
  number = {12},
  pages = {309},
  publisher = {The American Astronomical Society},
  doi = {10.3847/2515-5172/ad9dd4},
  url = {https://doi.org/10.3847/2515-5172/ad9dd4}
}
\bibliographystyle{aasjournal}

\end{document}